\documentclass[traditabstract]{aa}
\usepackage{txfonts}
\usepackage{natbib}
\usepackage{natbib}
\bibpunct{(}{)}{;}{a}{}{,} 
\usepackage{longtable}
\usepackage{multirow}
\usepackage{color}
\usepackage{graphicx}

\newcommand{\lya}{\ifmmode {\rm Ly}\alpha \else Ly$\alpha$\fi}
\usepackage{graphicx}
\usepackage{txfonts}
\usepackage{stackengine}
\DeclareUnicodeCharacter{2010}{-}
\begin{document}
\defcitealias{pentericci+14}{LP14}
\defcitealias{pentericci+11}{LP11}
\defcitealias{vanzella+11}{V11}
\title{CANDELSz7: a large spectroscopic survey of CANDELS galaxies in the reionization epoch}
\author{L. Pentericci\inst{1},
E. Vanzella\inst{2},
M. Castellano\inst{1},
A. Fontana\inst{1},
S. De Barros\inst{3},
A. Grazian\inst{1},
F. Marchi\inst{1},
M. Bradac\inst{4},
C. Conselice\inst{5},
S. Cristiani\inst{6},
M. Dickinson\inst{7},
S. Finkelstein\inst{8},
E. Giallongo\inst{1},
L. Guaita\inst{1,9},
A. Koekemoer\inst{10},
R. Maiolino\inst{11},
P. Santini\inst{1},
V. Tilvi\inst{12}
}
\institute{
INAF, Osservatorio Astronomico di Roma, via Frascati 33, I-00078 Monteporzio Catone, Italy \and
INAF, Osservatorio Astronomico di Bologna, via Gobetti  93/3 I-40129 Bologna,Italy  \and
Observatoire de Geneve, Universite de Geneve, 51 Ch. des Maillettes, 1290 Versoix, Switzerland \and
Department  of  Physics,  University  of  California,  Davis,  1
Shields Ave, Davis, CA 95616, USA \and 
School of Physics \& Astronomy, The University of Nottingham, University Park, Nottingham NG7 2RD, UK \and 
INAF-Osservatorio Astronomico di Trieste, via Tiepolo 11, I-34143 Trieste, Italy \and 
NOAO, 959 N. Cherry Ave, Tucson AZ 85719 USA \and
Department of Astronomy, The University of Texas at Austin, Austin, TX 78712, USA \and
Núcleo de Astronomía, Facultad de Ingeniería y Ciencias, Universidad Diego Portales, Av. Ejército 441, Santiago, Chile  \and 
Space Telescope Science Institute, 3700 San Martin Drive, Baltimore, MD 21218, USA \and
Kavli Institute for Cosmology, University of Cambridge, Madingley Road, Cambridge, CB3 0HA, UK 
\and
School of Earth \& Space Exploration, Arizona State University,
Tempe, AZ. 85287, USA
}
\abstract{
We present the  results  of CANDELSz7, an ESO large program aimed at 
confirming spectroscopically a homogeneous sample of z$\simeq$6 and z$\simeq$7 star forming  galaxies.
The candidates  were selected in the GOODS-South, UDS and COSMOS fields using  the official CANDELS catalogs based on  $H_{160}$-band detections. Standard color criteria, which were tailored depending on the ancillary multi-wavelength data available for each field,  were applied to select more than 160 candidate galaxies at z$\simeq 6$ and z$\simeq$ 7. 
Deep medium resolution  FORS2 spectroscopic  observations  were then conducted with  integration times ranging from 12 to 20 hours, to reach a Ly$\alpha$ flux  limit of approximately 1-3$\times 10^{-18}$ erg s$^{-1}$ cm$^{-2}$ at 3$\sigma$. 
For about 40\%\ of the galaxies  we could determine a spectroscopic redshift, mainly through the detection of a single emission line that we interpret as Ly$\alpha$ emission, or for some of the brightest objects ($H_{160}\leq 25.5$) from the presence  of faint continuum and sharp drop that we interpret as a Lyman break. 
In this  paper we present the  redshifts and main   properties of 65 newly  confirmed high redshift  galaxies. Adding previous proprietary and  archival data we assemble a sample of $\simeq 260$  galaxies that we use to explore the evolution of the Ly$\alpha$ fraction in Lyman break galaxies  and  the change in  the  shape of the emission line between $z\sim6$ and $z\sim7$. We also discuss  the accuracy of the CANDELS photometric redshifts in this redshift range.}
 
\authorrunning{L. Pentericci et al.} 
\titlerunning{CANDELSz7}
    
  \date{Received ; accepted}

   \maketitle
\section{Introduction}
The exploration of the reionization era is surely one  of the most challenging and fascinating tasks of present-day extra-galactic astronomy. 
For the first time we can compare precise  results from cosmic microwave background data from Planck  \citep{adam16} to observations of primeval galaxies, when the Universe was still largely neutral.
These observations help us to understand  the exact time-line of the reionization process, how it proceeded spatially, and which were the sources
that produced all or most of the ionizing photon budget.
The general consensus seems to be that galaxies, and in particular the faintest systems, were those providing most of the ionizing radiation \citep[e.g.,][]{bouwens+16b,fink15}, although faint AGN might also have played a role  \cite[e.g.,][]{giallongo+15}.
\\
To understand the evolution of the reionization process, one of the key quantities we would like to  measure is the fraction of neutral hydrogen present in the Universe and its evolution with cosmic time. In the future  SKA (and maybe its precursors)  will be able to detect directly the neutral hydrogen content in the early Universe by mapping  the 21 cm emission. In the meantime we must rely on  alternative  observational  probes, which allow us to  set indirect constrains on the  amount of neutral hydrogen. These include  deep optical spectra of high redshift QSO where we  can analyse  the Gunn-Peterson  optical depth (e.g. \citealt{fan2006}), the distribution of dark gaps (Chardin et al. 2018, McGreer et al. 2015), the analysis of damping absorption wings as in  Schroeder et al. (2013) and the analysis of GRB spectra \citep{totani+14}.  For Lyman break galaxies (LBGs) and Ly$\alpha$ emitters (LAEs) the most promising tools are 
 studying the prevalence of Ly$\alpha$ emission in star-forming galaxies (\citealt[][hereafter LP14]{pentericci+14},\citealt[][hereafter LP11]{pentericci+11},\citealt{ono+12,treu+13,schenker+14,caruana+14,tilvi+14}), the evolution of the clustering  and luminosity function  of LAEs \citep{ouchi+10,tilvi+10,sobacchi+15}.
\\
In particular, several groups have focused their attention on  the presence of the Ly$\alpha$ line in samples of LBGs. While from redshift $\sim$2 to $\sim$6 the fraction of galaxies showing a bright \lya\ emission line seems to be increasing steadily  \citep{stark+10,cassata+15}, there is a strong deficit of such lines in galaxies as we approach  z$\sim$7 (LP14; \citealt{tilvi+14,treu+13,ono+12,schenker+12};LP11;\citealt{fontana+10}). This is so far one of the strongest and perhaps the most solid evidence that at z$\sim$7 the Universe is partially neutral, since neutral hydrogen can easily suppress the visibility of the line. It would be much harder  to explain the observed drop in the Ly$\alpha$ fraction with a very rapid change in the physical properties of galaxies, such as the intrinsic dust content. The only alternative viable explanation is a sudden increase of the escape fraction of Lyman continuum photons \citep{mesinger15}; however  despite the recent progress in the discovery of real Lyman continuum emitters  (\citealt{shapley16,izotov+16b,vanzella+16}), this quantity remains  elusive  and assessing its evolution in the early Universe is extremely difficult.  
\\
The dramatic decrease of Ly$\alpha$ fraction at high redshift implies that the number of spectroscopically confirmed galaxies above z=6.5 is still  very low.  This emission line is at  present one of the few possible redshift indicators in the reionization epoch, although the Carbon line emission is becoming  a viable alternative, from transitions visible in the sub-mm (the  [CII] $158 \mu m$ line  e.g., \citealt{pentericci+16,bradac+17,smit17}), or in the UV domain (the CIII]1909\AA\ emission line,  \citealt{stark+17,lefevre17,maseda17,stark15}). All the above  results are  based on small data-sets, and  statistical fluctuations can be  very large. This is particularly true for the faintest LBGs, since most previous observations focused on the brighter candidates ($M_{UV}<-20.5$). 
The results also came from very   heterogeneous observational efforts, in terms of
wavelength coverage, sample detection  and selection, integration times etc: it is therefore hard to combine them and to assess their global statistical significance.
To overcome these problems, in 2013 we  started an ESO Large Program with FORS2 (Program ID 190.A-0685) to assemble observations of a much  larger and homogeneously selected sample of galaxies at z$\sim$6 and $\sim$7, to  place much firmer constraints on the decrease in the visibility of Ly$\alpha$ emission between these two epochs, and to assess if and how this decrease depends on galaxies' brightness. This is of particular relevance since the visibility of Ly$\alpha$-emitting galaxies during the Epoch of Reionization is controlled by both diffuse HI patches in large-scale bubble morphology and small-scale absorbers. Improved constraints on the relevant importance of these two regimes can be obtained by analyzing the full UV-luminosity dependent redshift evolution of the Ly$\alpha$ fraction of Lyman break galaxies \citep{kaki16}.

In this  paper we will present the observations and the results of our large program, while deferring a full analysis of the properties of the galaxies and on the constraints to reionization models to other papers.
In Section 2 we describe the target selection; in Section 3 the observations and data reduction procedure; in Section 4 we present the results and in Section 5 we discuss the observational properties of the detected galaxies. 
In a companion paper (De Barros et al. 2017),  we have discussed the physical properties of the redshift 6 sample, while in \citet{cast17} we have investigated the nature  of the GOODS-South  and UDS z$\sim$7 target, to probe  possible physical differences between \lya\ emitting and non-emitting sources.  Finally, in an upcoming work we will discuss more extensively the implication of our results for the reionization epoch.

We adopt a $\Lambda$-CDM cosmological model with $\mathrm{H}_\mathrm{0}=70$ km s$^{-1}$ Mpc$^{-1}$, $\Omega_\mathrm{m} = 0.3$ and $\Omega_\Lambda= 0.7$.  All magnitudes are expressed in the AB system \citep{okegunn83}.
\begin{figure}
\includegraphics[width = 9cm,clip=]{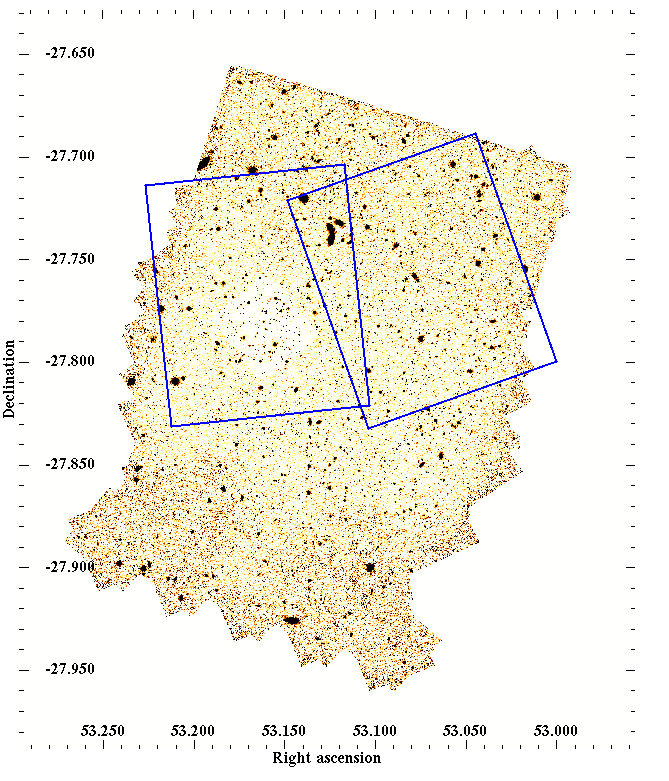}
\label{goodsfig}
\caption{The layout of the two FORS2 masks in the GOODS south field overplotted on the CANDELS $H_{160}$ image}
\end{figure}
\section{Target selection}
The targets were selected from the  CANDELS survey \citep{grogin+11,koekemoer+11} using the official catalogs in the UDS, GOODS-South and COSMOS fields
(\citealt{galametz+13,guo+13,nayyeri+17}, respectively for the three fields).  
The detection of the CANDELS catalog is  always based in the $H_{160}$-band which is the reddest  band available.  Most notably at redshift 6 and 7, 
this band is not affected by the presence of Ly$\alpha$ emission. This minimizes the possibility to bias part of the   sample towards strong line  emitters at any redshift. 
At variance with this, in the pre-CANDELS epoch, i-dropouts were typically selected from catalogs where the reddest detection band was the z-band \citep{stark+10,vanzella+08}.  This meant that  in the presence of a strong  line emission, fainter galaxies at high redshift were only detected because of  the brightening of the measured z$'$ magnitude due to the line flux,
and this was only partially compensated by the fact that galaxies in the lower part of the redshift selection window  would 
fall out of the selection (see a detailed discussion in  \citealt{stanway+07}) because their $i-z $ color would not be red enough to pass the cut.
In our case the  selection of galaxies in the $H_{160}$-band of course means that amongst objects with similar luminosity at 1500\AA\ rest-frame, those with a particularly blue slope could potentially fall out of detection.  However this bias is probably similar both at z$\sim$6 and 7 since the UV slopes of galaxies do not change appreciably between these two epochs in the magnitude range considered (e.g. \citealt{bouwens+14}). 
In addition to the CANDELS data, ($J_{125}$ and $H_{160}$ band plus $V_{606}$ and $I_{814}$ that are available for all three fields),  each of the field has multi-wavelength supporting observations both from ground instruments  and from space (see the above papers for a detailed list). 
In particular the GOODS-South field has the best optical multi-wavelength data, with deep HST imaging available in many different bands, and both UDS and GOODS-South 
have  supporting near-IR data (including deep HAWK-I K-band imaging) coming from the HUGS survey \citep{fontanaetal2014}. 
The different supporting data  result in slightly different selection criteria  in each field, although we attempted to apply selection criteria that were as uniform as possible for all fields.
The targets for our program were selected in the following way:
\\
1) For the $z\sim 7$  samples  we employed the color  criteria which were described in detail  in  \cite{grazian+12} separately  for the 
GOODS-South field and  the ERS sub-region (which has observations in the $Y_{098}$ filter instead of the $Y_{105}$,
therefore the criteria are slightly tailored to account for the difference in the transmission) and are respectively: 
 \begin{eqnarray*}
z-Y_{105}&>&0.8,\\
z-Y_{105}&>&0.9+0.75(Y_{105}-J_{125}),\\
z-Y_{105}&>&-1.1+4.0(Y_{105}-J_{125}),
\end{eqnarray*} 
for the GOODS-South field, and:
\begin{eqnarray*}
z-Y_{098}&>&1.1,\\
z-Y_{098}&>&0.55+1.25(Y_{098}-J_{125}),\\
z-Y_{098}&>&-0.5+2.0(Y_{098}-J_{125}),
\end{eqnarray*} 
for the ERS area.
For the non-detection in photometric bands bluer than
$Z$, we adopt the same criteria used in \citet{castellano+10a,castellano+10b}
and in \cite{grazian+12} which are  $S/N<2$ in all BVI HST bands and $S/N<1$ in at least two of them.
\\
For the UDS and COSMOS  fields, where the only photometry available
from space is in $V_{606}$, $I_{814}$, $J_{125}$, and $H_{160}$ bands,
we adopt an  $I_{814}$-dropout color described also in \citet{grazian+12}, which gives a more extended redshift window for selecting galaxy candidates ($6.4<z<8.5$) and is the following: 
\begin{eqnarray*}
I_{814}-J_{125}&>&2.0,\\
I_{814}-J_{125}&>&1.4+2.5(J_{125}-H_{160})
\end{eqnarray*}
 In this case a stricter non-detection was required in the only HST band bluer of the Lyman break available, i.e., $S/N(V_{606})<1.5$, and non-detections ($S/N<2$)  in all ground-based images bluer of the R-band.  
 The redshift selection functions  for these different color criteria adopted are shown in Figure 1 of \cite{grazian+12}.
\\
2) For the $z\sim 6$  sample in the GOODS-South field we have employed a selection that is similar (but not identical) to the  one used in \cite{bouwens+15b}:$
i_{775}-z_{850}>1.0 \land Y_{105}-H_{160}<0.5$,  with a requirement for a non-detection in either  the  B or V band as 
$(S/N(B_{435})<2\land V_{606}-z_{850}>2.7)\lor S/N(V_{606})<2$.
\\
For the COSMOS and UDS fields we used the following criteria:  
$(i-z)>1.0 \land (J_{125}-H_{160}) <0.5 $ with a requirement for non-detection in both the U and B band as   $S/N(UB)<2$.
\\
3) In addition  we also selected galaxies that were not compliant with the above color criteria, but which had a photometric redshift between 5.5 and 7.3, i.e., the approximate range  where we can expect to detect the Ly$\alpha$ emission with the adopted observational set-up. 
The selection with photometric redshift was done to complement 1) and 2) above, since  it can recover objects that are scattered out of the color criteria, e.g., because of uncertain photometry or because of the presence of the \lya\ emission line.
\\
The photometric redshifts adopted were the official CANDELS best redshifts   presented in \citet{santini+2015} for the GOODS-South and UDS fields, and by \citet{nayyeri+17} for the COSMOS field. Briefly, they are  based on a hierarchical Bayesian approach that combines the full probability distribution functions PDF(z)  of individual redshift determination provided  by several different CANDELS photo-z investigators.  
The precise techniques adopted to derive the official CANDELS photometric redshifts, as well as the individual values from the various participants, are described in details by \cite{dahlenetal2013} and in the above papers.

 Finally all candidates selected in the above categories were visually inspected in all photometric bands available  to remove possible false detections due to e.g., residual defects in regions of poorer imaging  quality. The blue bands were also smoothed to ensure that the non-detection  was solid.  In addition galaxies  which could plausibly be at high redshift but which had very close objects (angular separation $\leq 1-2''$) that could hamper their spectroscopic identification were also removed for the samples.
\\
\begin{table*}
\caption{Masks characteristics}
\begin{center}
\begin{tabular}{ccccccccc}
Mask & RA & DEC &  $T_{exp}$ & z$\sim$6 & z$\sim$7 & AGN & others & Ntot \\
      & J2000 & J2000 & hours & \\
\hline
UDS1    & 2:17:29.6 &-05:57:22.4 & 12.2 &  6  & 9  & 11 & 5 & 31 \\     
UDS2    &  2:17:59.9  & -05:12:52.1    & 12.2 &  11 & 6  &  8 & 3 & 28    \\
UDS3    & 2:17:04.5   & -05:11:03.3    & 12.2 &   5 & 11 & 3  &11 & 30    \\
COSMOS1 & 10:00:30.7 & +02:16:56.9      & 12.2 &  12 & 10 & 7  & 3 & 32      \\
COSMOS2 &  10:00:37.4 &+02:27:10.7   & 12.2 &  5  & 12 & 5  & 3 &  25  \\
COSMOS3 &  10:00:22.5 & +02:23:52.4  & 12.2 &  10 & 8  & 5  & 5 & 28    \\
GOODS-S1&  03:32:17.9  &-27:45:54.4     & 20   & 18  & 18 & 3  & 1 & 40 \\
GOODS-S2& 03:32:39.6   &-27:46:02.8     & 20   & 12  & 15 & 2  & 2 & 31\\
\hline
TOTAL   &    &     &       & 78 & 89  & 44  & 33  & 245
\end{tabular}
\end{center}
\label{tab:fields}
\end{table*}
\\
We designed  three  masks for each of the COSMOS and UDS fields and two masks for  GOODS-South. 
The  masks were designed  to allocate the largest number of z$\sim$7  galaxies as first priority, and then in order of decreasing priority, z$\sim$6  candidates, AGN candidates (i.e., X-ray and Herschel sources) and other fillers. The mask construction was driven only by  geometrical constraint without any other bias.  In each mask we also reserved a number of slits (two per each of the  FORS2 chips) to observe bright objects: this is required since  our objects (with the exception of some AGN and fillers) were not visible in individual 20 minutes exposures and having bright objects was usefully to check centering and trace the spectra.  
Typically each mask contained between  30 and 40  useful targets. 
The total number of objects placed in each mask divided by category is reported in Table 1. 
In Figures 1,2 and 3 we show the position of all FORS2 masks over-laid on the CANDELS $H_{160}$-band images for the GOODS-South, UDS and COSMOS respectively.
In  Figure \ref{fig:colorfig} we show the color-color selection diagrams for all the z$\sim$7 candidates inserted in the masks, for the different fields and areas. The  density of observed candidates at z$\sim$7 is similar for the UDS and COSMOS fields ($\sim0.20-0.22/arcmin^2$) while it is higher for GOODS-South (0.42/arcmin$^2$) because of the increased depth of the CANDELS detection catalog.
\begin{figure*}
\includegraphics[width = 14cm,clip=]{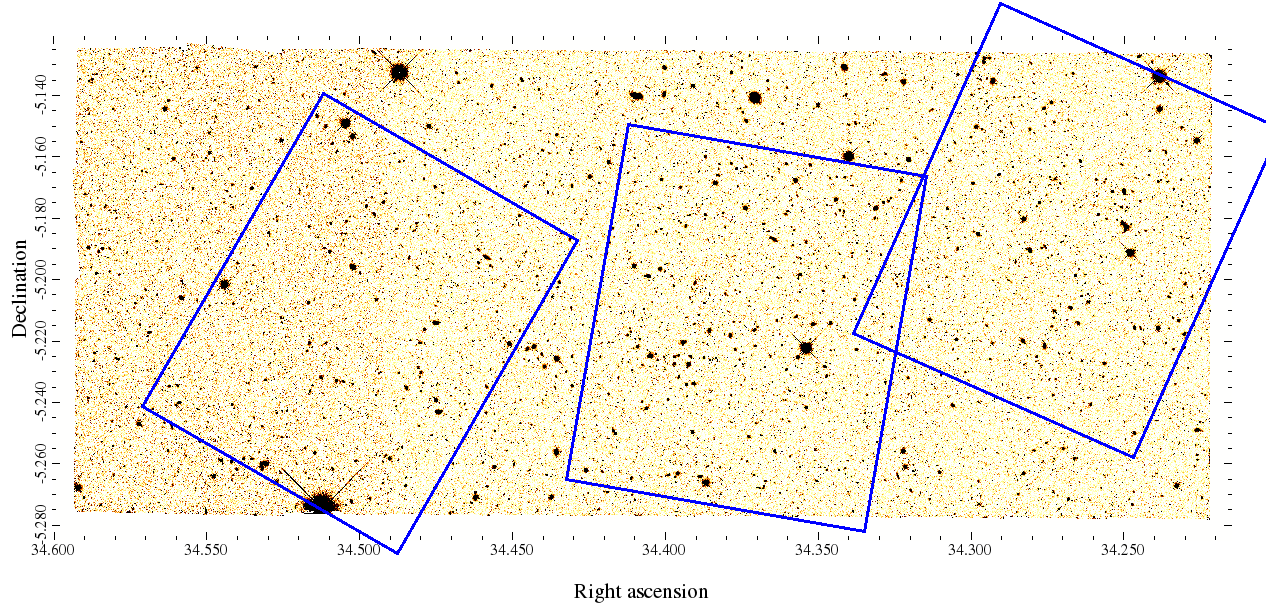}
\caption{The layout of the three FORS2 masks in the UDS field overplotted on the CANDELS $H_{160}$ image}
\label{udsfig}
\end{figure*}

\begin{figure}
\includegraphics[width = 8cm,clip=]{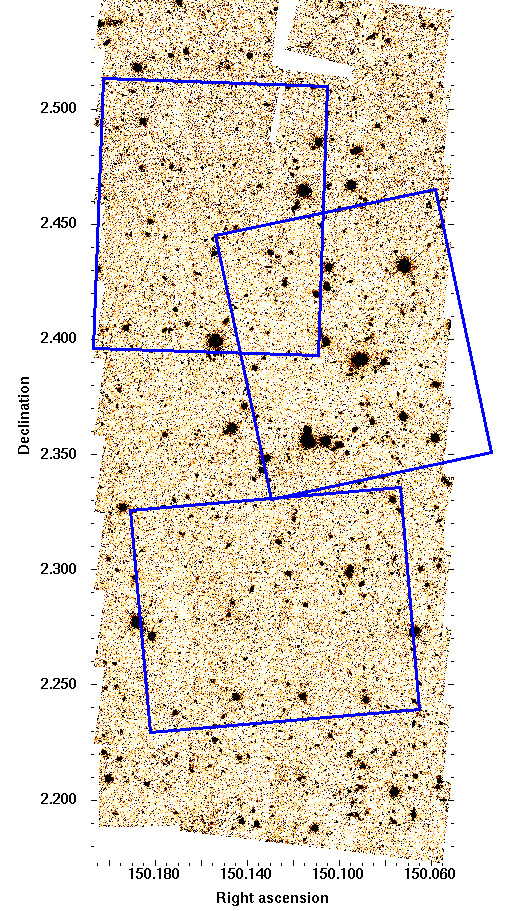}
\caption{The layout of the three FORS2 masks in the COSMOS field overplotted on the CANDELS $H_{160}$-band image }
\label{udsfig}
\end{figure}
\begin{figure*}
\includegraphics[width = 9cm,clip=]{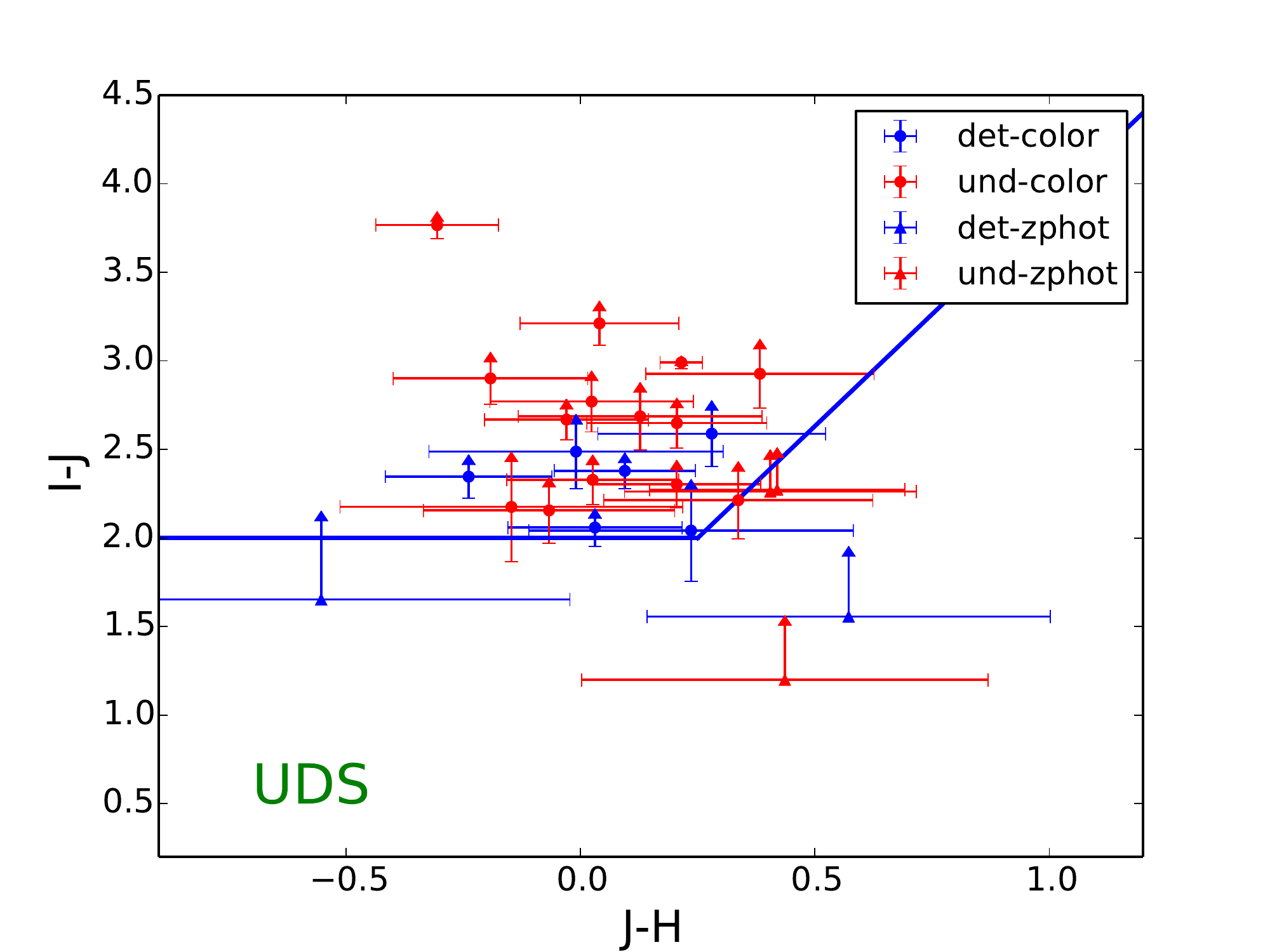}
\includegraphics[width =9cm,clip=]{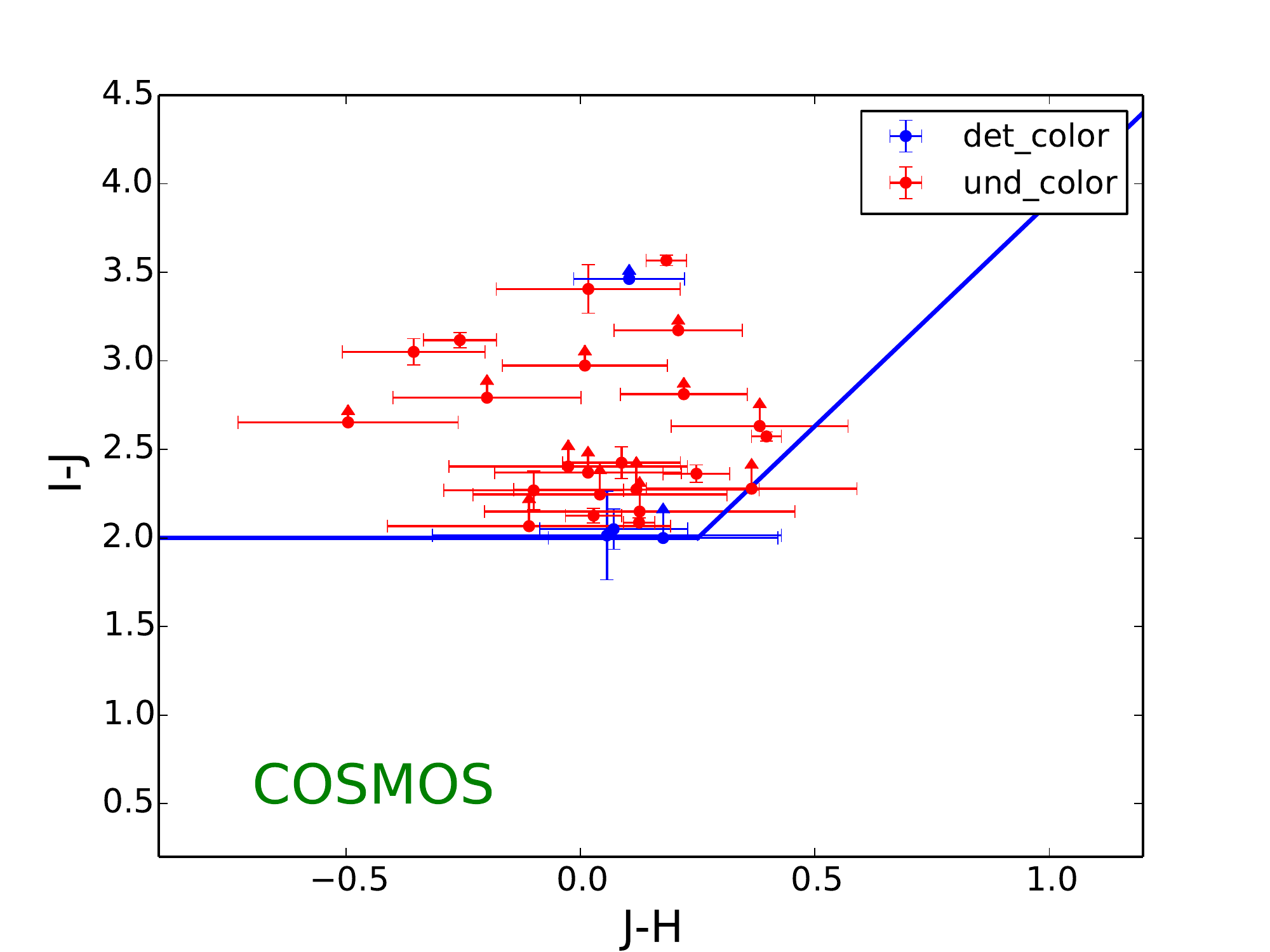}
\includegraphics[width = 9cm,clip=]{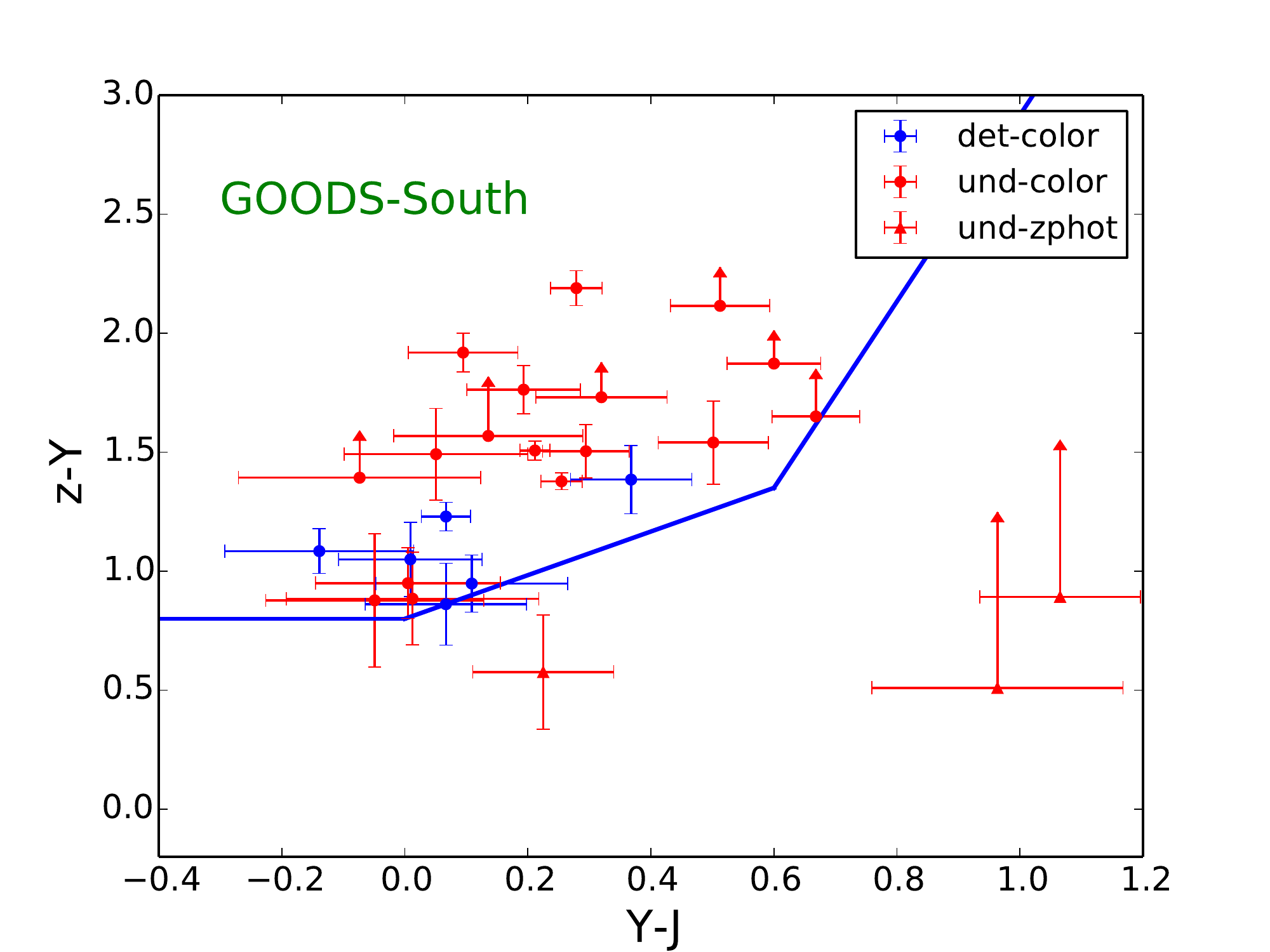}
\includegraphics[width =9cm,clip=]{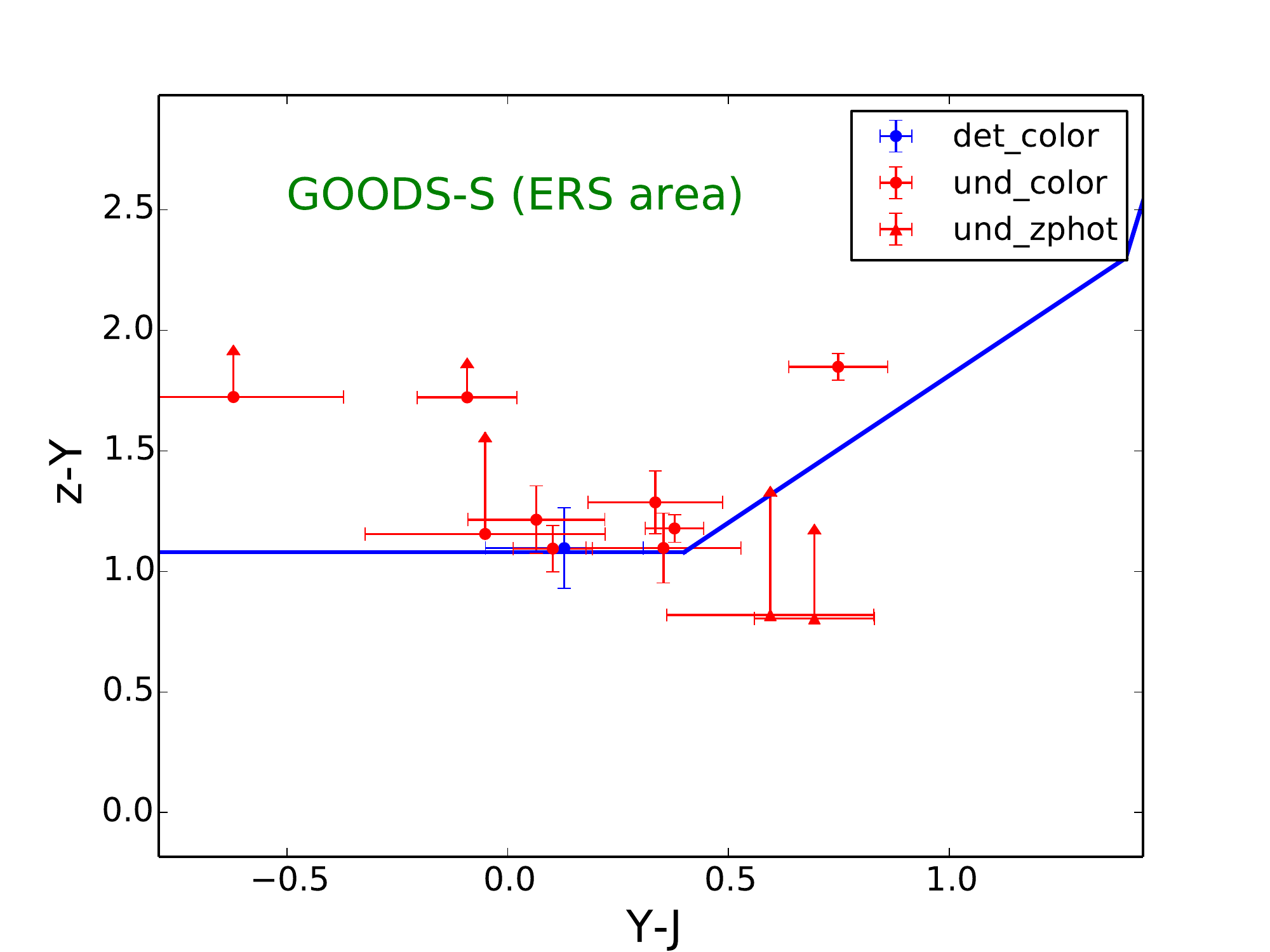}
\caption{The color-color selection diagrams in the different regions observed in this work: upper panels show  the $J_{125}-H_{160}$ vs $I-J_{125}$ diagram for the UDS field (left) the and COSMOS field  (right). Bottom panels show the  $Z_{850} - Y_{105}$ vs. $Y_{105} - J_{125}$ diagram for the GOODS-South field (left) and the $Z_{850} - Y_{095}$ vs. $Y_{095} - J_{125}$ for the  ERS area (right). The blue circles are the detected objects, and the right circles the undetected ones. Blue and red triangles represent respectively confirmed and unconfirmed objects that were selected only by means of their photometric redshift. The solid lines are the color cuts representing the different  criteria described in Section 2. }
\label{fig:colorfig}
\end{figure*}

\section{Observations and data reduction}
Observations were taken  with the FORS2 spectrograph on 
the ESO Very Large Telescope. We used the 600Z holographic grating, which provides the highest
sensitivity in the range $8000-10000$\AA\  with a spectral resolution
 $R\simeq 1390$ and a sampling of 1.6\AA\  per pixel for a slit width of 1$''$. The total slit length varied in order to allocate as many objects as possible but still allow for an accurate sky subtraction. Compared to previous  observations where the length was always kept longer than 10-12$"$ (e.g., LP11) in some cases our slits were as short as 8$''$. The objects were placed at the center of the slit whenever allowed by geometrical constraints  (approximately in 90\% of the cases) and in all cases at least 4$''$ from the border. 
 \\
The observation strategy was almost identical to the one adopted in LP11 and  LP14:  series of spectra with 1200 s integration (instead of the 665 s  that were used in previous observations) were taken at two different positions, offset by $4"$ (16 pixels) in the direction perpendicular to the dispersion, with a pattern ABBA.  
The total net integration time was  approximately 12 hours for the masks in the UDS and COSMOS fields, and 20 hours for the masks in the GOODS-South field
(the precise values for each mask are reported in Table 1).
Observations were carried out in service mode and we requested a seeing limit of 0.8$''$,  clear conditions and moon illumination below 0.3.  The observations were spread over 6  semesters, from January   2013 to January   2016.

Data were reduced using the well-tested pipeline developed by \cite{vanzella+08} and described at length in \cite{vanzella+14}, specifically tailored for the reduction of very faint spectra, which we already used for all  our previous observations of z$\sim$7 galaxies (V11, LP11, LP14). Briefly 
after a standard flat-fielding and  bias subtraction,  the sky background was  subtracted between consecutive exposures, exploiting the fact that the target spectrum
is offset due to dithering.  An alternative sky-subtraction method, consisting in   
fitting a polynomial function to the background was also applied, giving similar or slightly worse results.
A standard wavelength  calibration was performed using arc lamps (He, Ar) that provide sharp emission lines over the entire spectral range observed. Before combining frames, particular care  has  been  devoted  to  the  possible  offset  along  the  wavelength  direction,  by  measuring  the  centroids  of  the  sky  lines in the wavelength interval 9000–9900 \AA. Systematic translations of the wavelength scales were corrected. 
The 1-dimensional spectra were extracted using apertures which encompassed all the Ly$\alpha$ flux based on individual visual inspection. Finally, the 1-d spectra were flux-calibrated using observations of spectro-photometric standard stars. Based on the analysis of the
standard stars observed with the same setup of science targets,
we derive that the relative error due to flux calibration is less
than 10\%. 

 We expect slit losses to be small, given the extremely
compact size of the targets, the very good image quality
during the observations and the extremely careful centering procedure during the production of the masks. 
To give an estimate of the possible effect, we refer to the simulation by \cite{lemaux}, who showed how the slit losses depend both on image quality and on the size of the objects.
We retrieved the sizes of the galaxies $r_e$, measured in the $J_{125}$ band using Galfit \citep{vdw}. The  median $r_e$ for our detected objects is 0.148$''$, in  agreement with the average size of z$\sim$7 galaxies determined by \cite{grazian+12} also in the  $J_{125}$ band. As previously specified, the upper limit for the seeing requested by the  survey was 0.8$''$ and in 80\% of the cases our OBs were graded A, implying  that all  constraints were  respected (with seeing between  0.4$''$ and 0.8$''$). Only for the remaining 20\% of the OBs the seeing was between 0.8$''$ and 1.0$''$. 
According to the simulation of \cite{lemaux} (Figure 3 of that paper), for these conditions and target sizes, slit losses should be about 15\%.
 Since all masks contained a random  mixture of z$\sim$6 and z$\sim$7 sources, the average seeing conditions  are the same for the two samples, and the sizes do not change sensibly between these two redshifts. In conclusion we do not take into account possible slit losses  in the subsequent discussion. However we are aware that this could be not true in all cases, as recently shown by the  discrepancy found  between the fluxes measured by HST grism  and ground based slit-spectroscopy in some high redshift galaxies \citep{huang+16,tilvi16,hoag}. 

In some cases the sky-subtraction was not optimal, e.g., when the slit was particularly short, or there was a  lower redshift interloper falling within the same slit, a very bright object close to the border,  or  the target itself  was close to the border of the slit. For these reasons we could not detect or set significant  limits for about 7-10\% of the objects observed. 

In this paper we also report new results from archival observations  that were taken  as part of the  program ESO 088.A-1013 (PI Bunker). This program employed the same observational setup used by our large program, with a total net integration time of 27 hr on a single mask. It observed a mixture of i- and z-dropouts in the Hubble Ultra Deep field. The description and some of the results  were presented by \citet{caruana+14}. We applied the same color criteria described above to their targets (the target list was derived using the information contained in the headers), and re-reduced the data in the relevant slits with our own pipeline, which is particularly tailored to the detection of  faint emission lines. We could confirm several additional high  redshift sources, which were also in our selection catalogs,  with the most distant one already reported in our previous work (CANDELS 34271 at z=6.65  LP14).  

\section{Results} 
\begin{figure}
\includegraphics[width = 9cm,clip=]{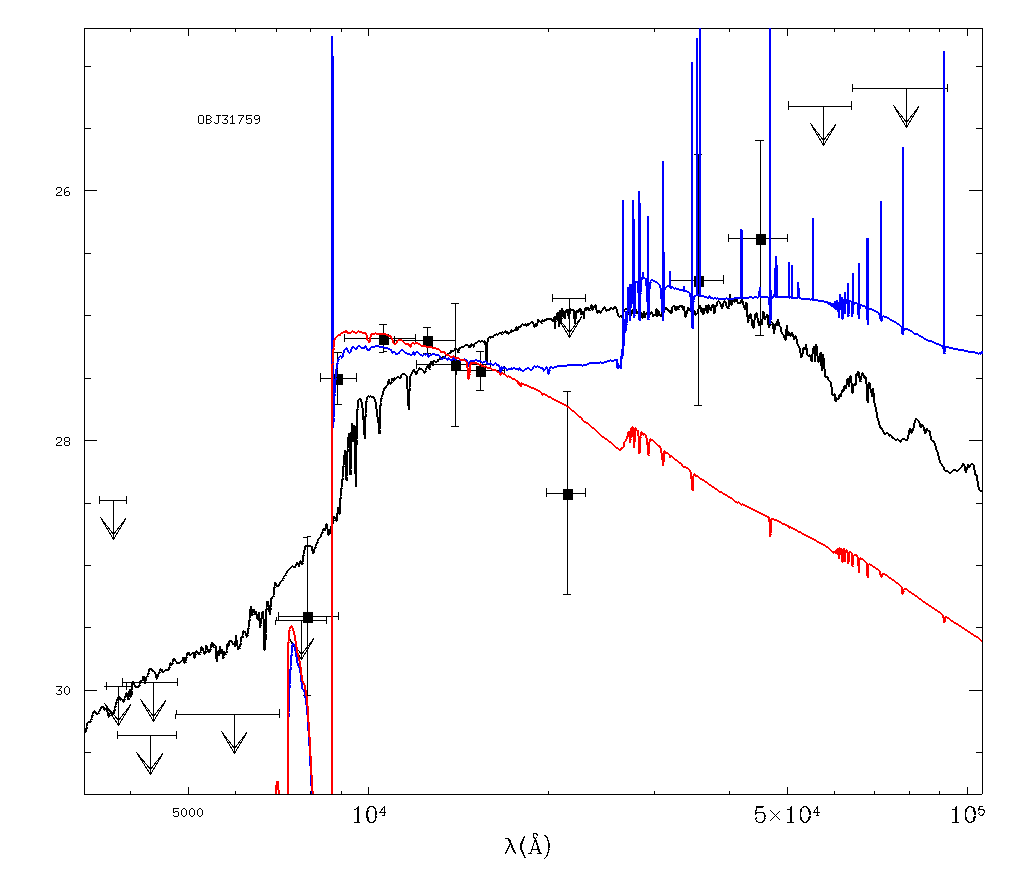}
\caption{The observed photometry of object GOODS-S 31759 with the  best fit solutions at the two possible spectroscopic redshift of z=1.393 (black curve) and z=6.34 (red curve and blue curve) corresponding respectively to the identification of the emission line visible in the spectrum  as the [OII] doublet or the \lya\ line. The two SEDs for the high redshift solution correspond to models with and without nebular contribution (respectively blue and red). }
\label{fig:sed}
\end{figure}

We searched for spectral features by an automatic scanning of the 2-dimensional spectra complemented by visual inspections by LP and EV. We report only features that  were confirmed independently by LP and EV. Before validating a detection we also ensured that: 1)  the position of the putative line in the 2-dimensional  frame is fully consistent  with the position of the target along  the slit in the mask; 2) in none of the individual 2-d frames there are spikes or artifacts at the position of the line; 3) for features with $S/N \geq 5$ a further confirmation of the reality of the feature must be the presence of negative residuals at positions $\pm 16 pixels$ along the Y-axis in the 2-d frame.
In the spectra of  56  objects we detect a single emission line that we interpret as Ly$\alpha$   (see the detailed  discussion below).  In one further  case (COSMOS ID 8474) we detect two emission lines that are consistent with H$\beta$ and [OIII] from a lower redshift galaxy at z=0.661. This object had been selected both by the i-dropout criteria and for its  photometric redshift.  In one further galaxy,  GOODS-S ID 31759 we detect two emission lines that are separated by 6 \AA\ and  are consistent with the [OII] emission doublet (3726/3729\AA) from a galaxy at z=1.393. This would be  a low luminosity [OII] emitter with Log(L([OII]))=41.2.  However there is a strong sky-line exactly between the two components: therefore it could be also possible that the line is actually  \lya\ at z=6.34, which appears split into two components because of the strong sky-line. In this latter case the line would be intrinsically very bright since the measured flux (with no correction) is already   1.7$\times 10^{-17}$ erg s$^{-1}$ cm$^{-2}$. The broad band photometry of this object actually strongly favors the  high redshift solution, with the photometric redshift constrained between 6.02 and 6.33 at 68\% confidence. In Figure \ref{fig:sed}   we show the broad-band photometry of the galaxy together with the best fitting SED, respectively at z=1.393 and z=6.34:  the high redshift solution  better matches the photometry, especially if we include the contribution from  nebular emission   (blue line).

In all other cases, for single emission line spectra, alternative identifications have been discarded with good confidence for the following reasons: 1) the  wavelength  range  covered by our spectra is large enough that if the detected lines were  H$\alpha$, H$\beta$  or either of the two [OIII] 5007,4959 \AA\  components, there would be at least another line visible  in the same spectrum, unless the object had very anomalous line ratios; 2) a possible identification with [OII], which could be naturally consistent in some targets with photometric redshift around 1.2-1.3, is also not likely since  we would resolve  the doublet at 3727-3729\AA,  as is the case for the target  described above. In \citet{vanzella+11} we showed other  examples of such low redshift doublets clearly resolved in the observations (Figure 2 of that paper). Note that this is only partially true in a few cases where the detected line falls very close  to a sky-line and the second component of the doublet could fall on top of the sky-line; 3)  when the signal to noise is high enough (17 targets), a sharp asymmetry of the line is  observed. This shape  is typical only of Ly$\alpha$ at high redshift and it is not observed for any other emission line; 4)  for the brightest objects (approximately H$ \leq 25.5$)  we also observe a weak continuum red-ward of the line but not blue-wards of it, which would not be observed for lower redshift objects and is perfectly consistent with  a large drop observed between the z and Y  (or i and z) broad band photometry.
The redshift are determined from a fit of the \lya\ line peak. The errors,  estimated following \cite{lenz} depend on the resolution and on the S/N of the line and are typically $\sim$0.002 for the faintest line emitters (S/N$\sim$5) and $<0.001$ for the brightest sources (S/N >10).
\\
In the spectra of nine further galaxies, typically our brightest candidates around $z\sim6$,  we detect only weak continuum with a  sharp break but no emission line.    We interpret this  discontinuity  as the Ly$\alpha$ forest  break. To estimate the wavelength of the break we smooth  both the 1d and 2d  spectra and we  determine the wavelength at which the flux becomes consistent with zero (i.e. the noise)   by  searching for a change in slope in the  cumulative sum of the flux (e.g., \citealt{watson}).  These redshifts are obviously less accurate then those based on the \lya\ emission line. The uncertainty is derived by changing the smoothing parameter and repeating the above measurement and it is $\pm0.1$.  In these cases we have  further confidence that  the redshift assignment is correct given the good agreement between the photometric redshift and the spectroscopic one, hence excluding with high probability that the break in the continuum is tracing the 4000\AA\ break, potentially associated with a lower redshift solution.

Taking into account the objects which had problems in the data reduction process (as detailed in Section 3), the overall success rate for redshift determination of the Large Program for  high redshift targets was $\sim$40\%. In Figures \ref{fig:specUDS}, \ref{fig:specGOODS} and \ref{fig:specCOSMOS} we show the 2-dimensional spectra of all the confirmed galaxies.

\begin{figure}
\includegraphics[width = 9cm,clip=]{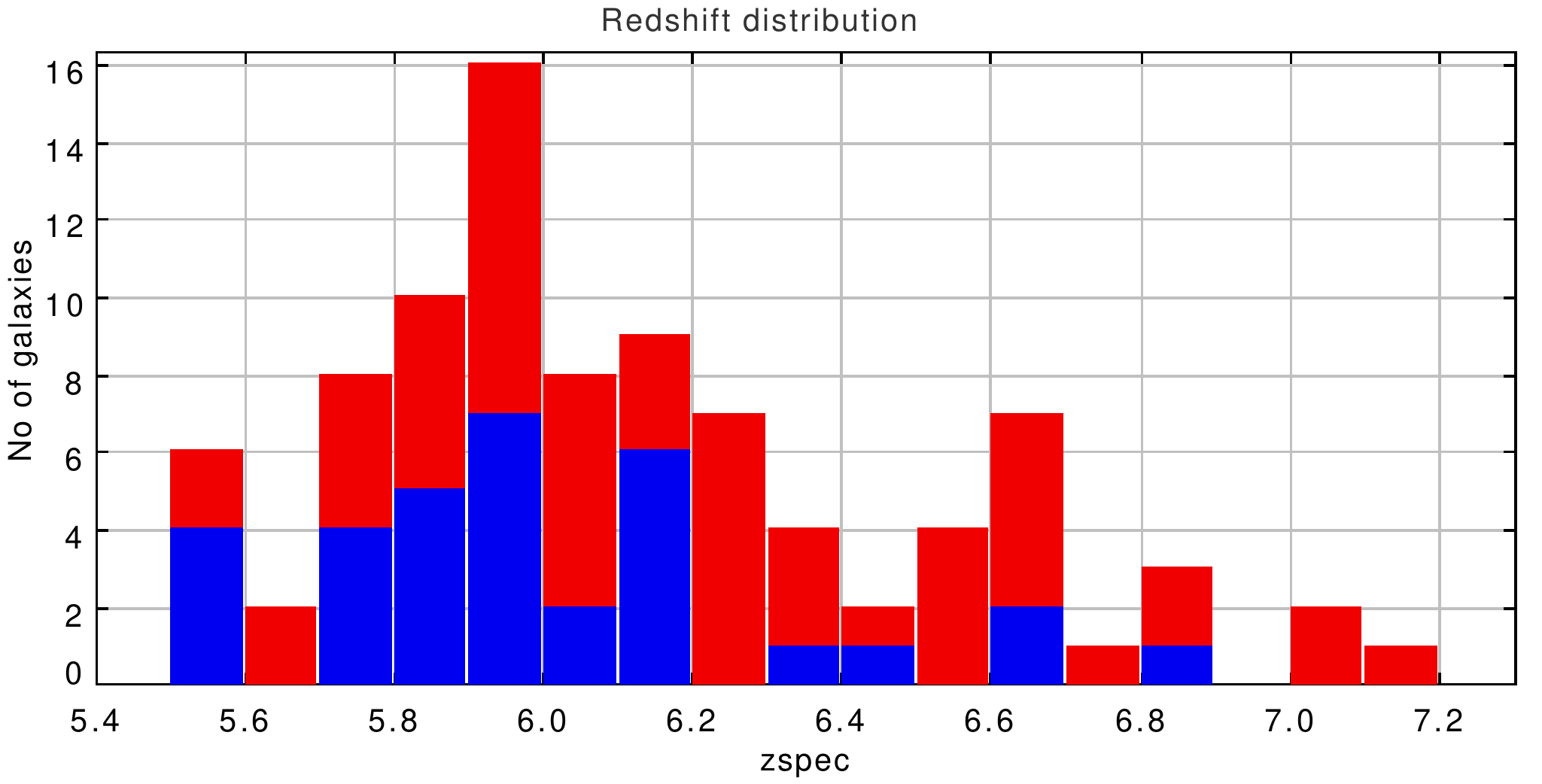}
\caption{The spectroscopic redshift distribution for all confirmed galaxies in the three  CANDELS fields analyzed  in this work between z=5.5 and 7.2. In blue we show previous redshifts in these  fields from the literature  (\citealt{pentericci+14,caruana+14,curtislakeetal2012,pentericci+11,fontana+10,vanzella+08}), while in red we show the new redshifts from this work.}
\label{fig:redshift}
\end{figure}

\begin{figure*}
\includegraphics[width = 9.5cm,clip=]{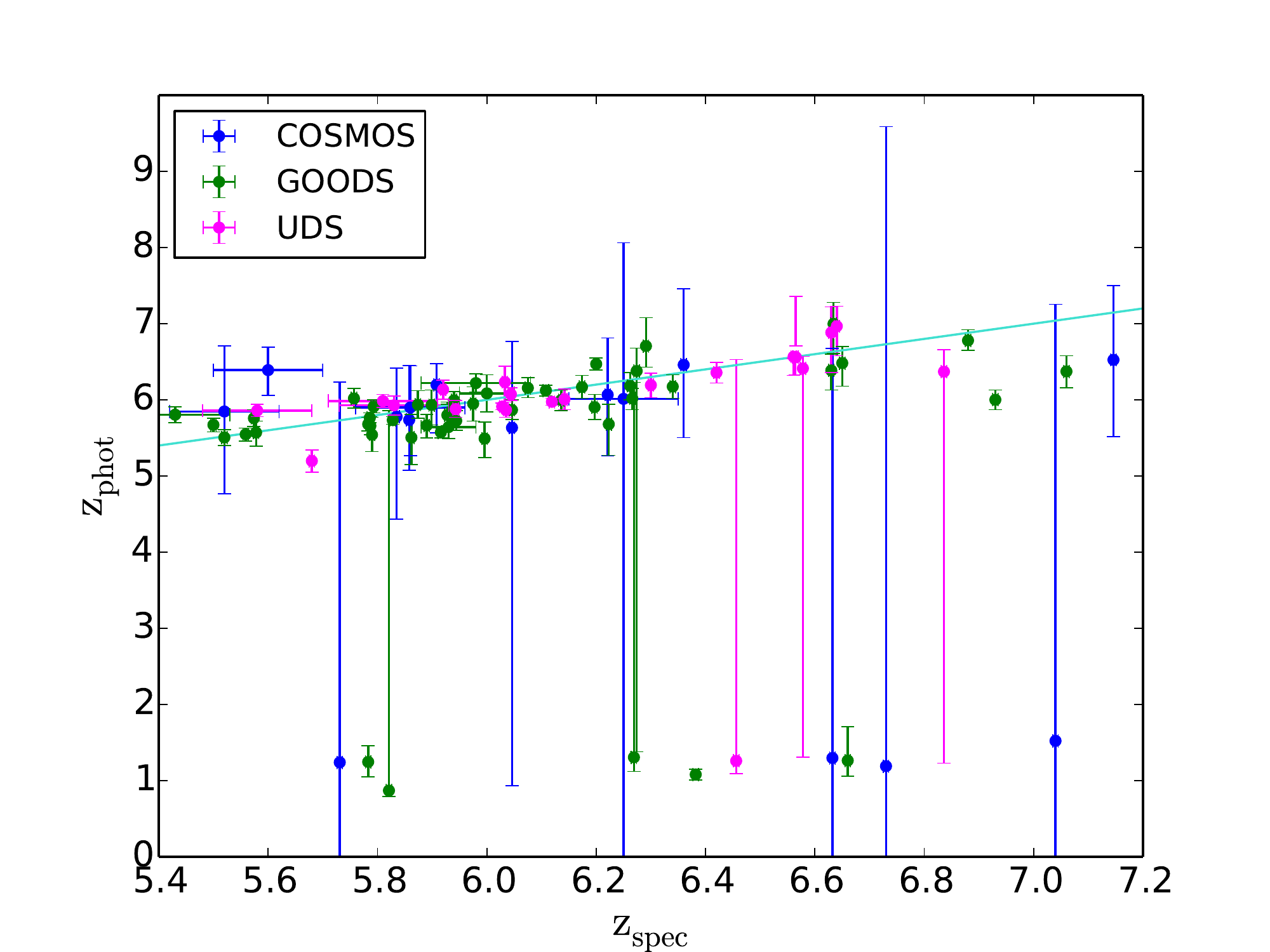}
\includegraphics[width = 9.5cm,clip=]{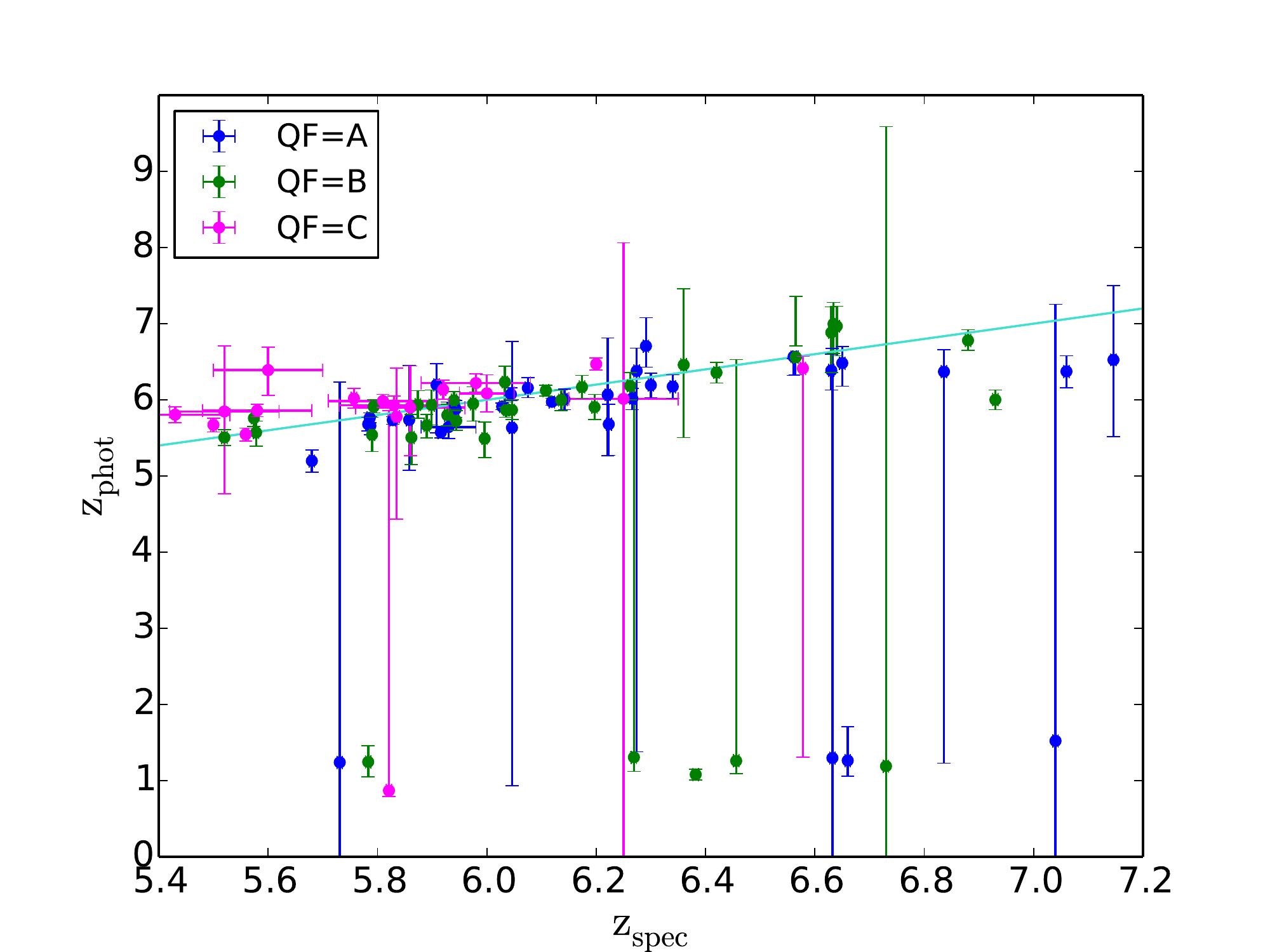}
\caption{The spectroscopic  redshift vs CANDELS photometric redshifts for the objects presented in this paper, left  represented with  with different color codes for  the three different fields (COSMOS, GOODS-South  and UDS in blue, green and red respectively), and right with different color codes for the different quality flags (flag A, B and C as blue, green,red respectively). The error bars represent the 68\% upper and lower uncertainties of the photo-z (see \cite{dahlenetal2013} for details).  }
\label{fig:specz}
\end{figure*}
We assigned a Quality Flag (QF) to each spectrum, indicating the reliability of the redshift  identification  which is mainly due to the S/N of the Ly$\alpha$ line: A is a completely secure identification (when we observe a clear asymmetry of the  line and in some cases the  continuum red-ward of it) and C is the most uncertain, typically assigned to those objects where we only observe a  weak continuum and a  break. A similar redshift flag scheme was adopted previously by \citet{vanzella+08} for the GOODS south spectroscopic campaign. The flags were first  independently assigned by LP and EV and then an agreement was reached in  case of initial discrepancies.
\\
In Tables 3, 4 and 5  we report the redshift identifications for a total of 67  objects, 21, 18 and 28 respectively in the UDS, COSMOS and GOODS-South fields. These are the confirmed galaxies at z$\sim$6 and 7  plus the two lower redshift  interlopers discussed above; we will report on AGN and other filler galaxies in future works.
For each object we report the CANDELS ID, the RA and Dec, the CANDELS $H_{160}$-band magnitude, the official CANDELS photometric redshift (\citealt{santini+2015} for the GOODS-South  and UDS fields, \citealt{nayyeri+17} for the COSMOS field),  the effective radius $r_e$  in the $J_{125}$ band  \citep{vdw}, and the spectroscopic redshift  with its quality flag. For the objects with an identified Ly$\alpha$ emission line,  we report the total flux measured from the 1-d spectra, and the Ly$\alpha$ rest-frame EW (REW). This last  quantity  is determined by estimating the continuum  emission at $1300\times (1+z)$  \AA\  from the available  broad band photometry, extrapolating from the nearest filter using a power law with the appropriate $\beta$ slope (see below).  We remark that we do not apply any slit loss correction to the Ly$\alpha$ flux:  our objects are mostly very compact and the slit losses should be minimal, however the reported REW might be in some cases underestimated, particularly for the few objects with large sizes.  The fluxes of the Ly$\alpha$ lines vary between $1.5\times 10^{-18}$ to $2\times10^{-17}$ erg s$^{-1}$ cm$^{-2}$
and the REW  span a range from 3 to 110 \AA. For the objects with no Ly$\alpha$ emission we report a limit (3$\sigma$) on the REW that is derived using accurate simulations of the reduction process presented in \cite{pentericci+14} and \cite{vanzellaetal2014} and assuming that the undetected i-dropouts are exactly at $z=6$ and the undetected z-dropouts are at $z=6.9$. 
\\
In the Tables we also list the absolute UV magnitude,  $M_{UV}$ and the  slope $\beta$ of the UV continuum. These  parameters are  obtained  by the  common power-law approximation for the UV spectral range $F_\lambda \propto  \lambda^\beta$, and estimated by fitting a linear relation through the observed AB magnitudes of the objects, excluding the band than contains the Ly$\alpha$ emission line, i.e., the z-band for objects around z$\sim$6 and the Y-band for higher redshift objects. 
We use broad band  fluxes measured in 2$\times$FWHM apertures instead of the isophotal photometry to estimate the UV slope and $M_{UV}$,  after verifying that this choice improves the  stability of the log(F) vs log($\lambda$) fit \citep{castellano+12}, compared to simply using the CANDELS isophotal magnitudes.
Finally in Tables 3, 4 and 5 we also indicate the selection criteria for each target, 1 is the z-dropout color, 2 is the i-dropout color  and 3 is the photometric redshift.
The distribution of the new redshift identification is presented in Figure \ref{fig:redshift}, together with previously known spectroscopic redshift in the three CANDELS fields from previous works \citep{caruana+14,pentericci+14,curtislakeetal2012,pentericci+11,fontana+10,vanzella+08} in the same redshift range.

The data on the high redshift objects will be released through the ESO science archive facility. We plan to release both the 2-dimensional spectra presented in Figures \ref{fig:specUDS}, \ref{fig:specCOSMOS} and \ref{fig:specGOODS} as well as the 1-dimensional extracted spectra, with associated noise spectra.

\section{Properties of confirmed galaxies}
\subsection{Accuracy of photometric redshifts}
\begin{figure}
\includegraphics[width = 9.5cm,clip=]{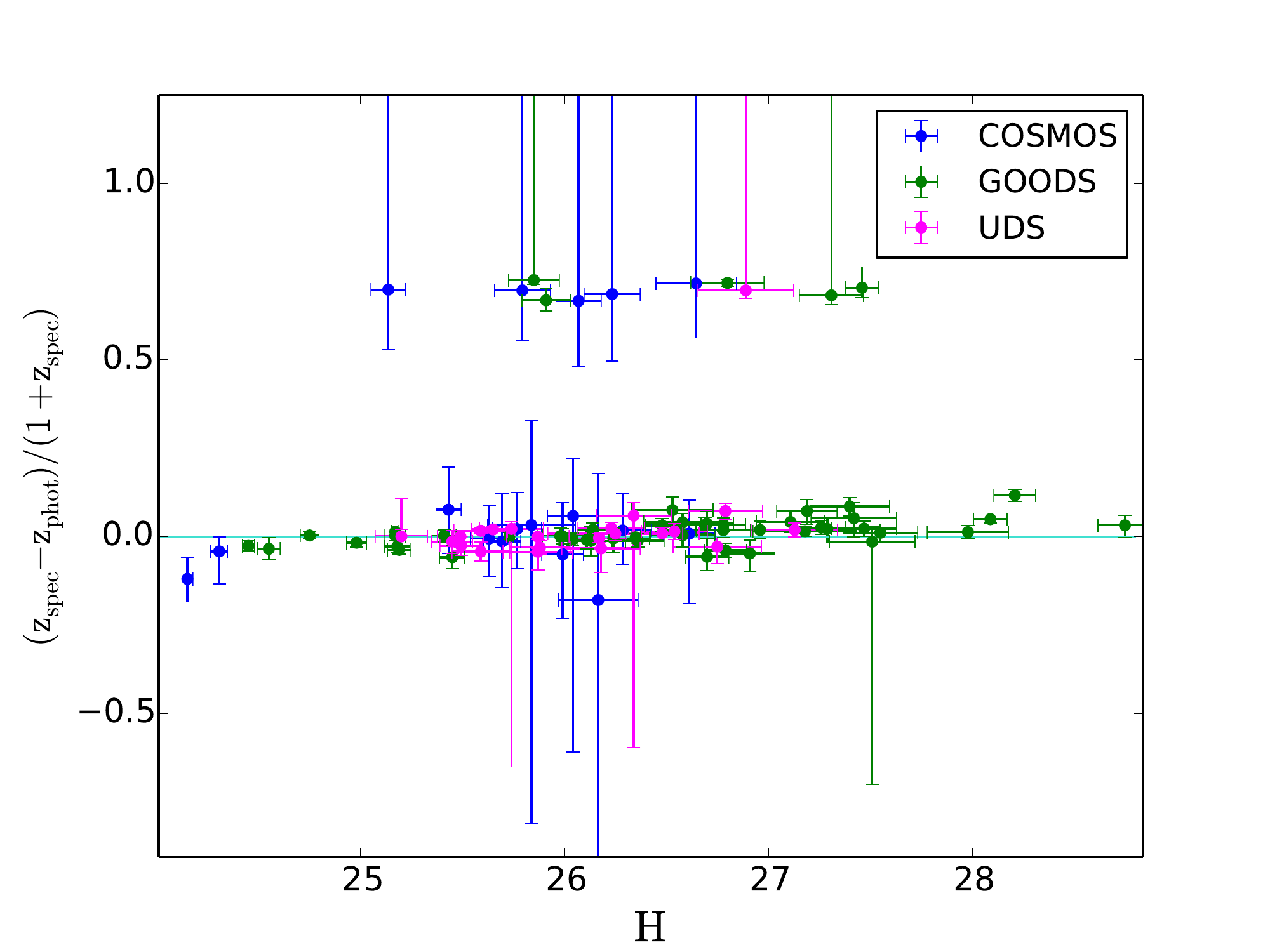}
\includegraphics[width = 9.5cm,clip=]{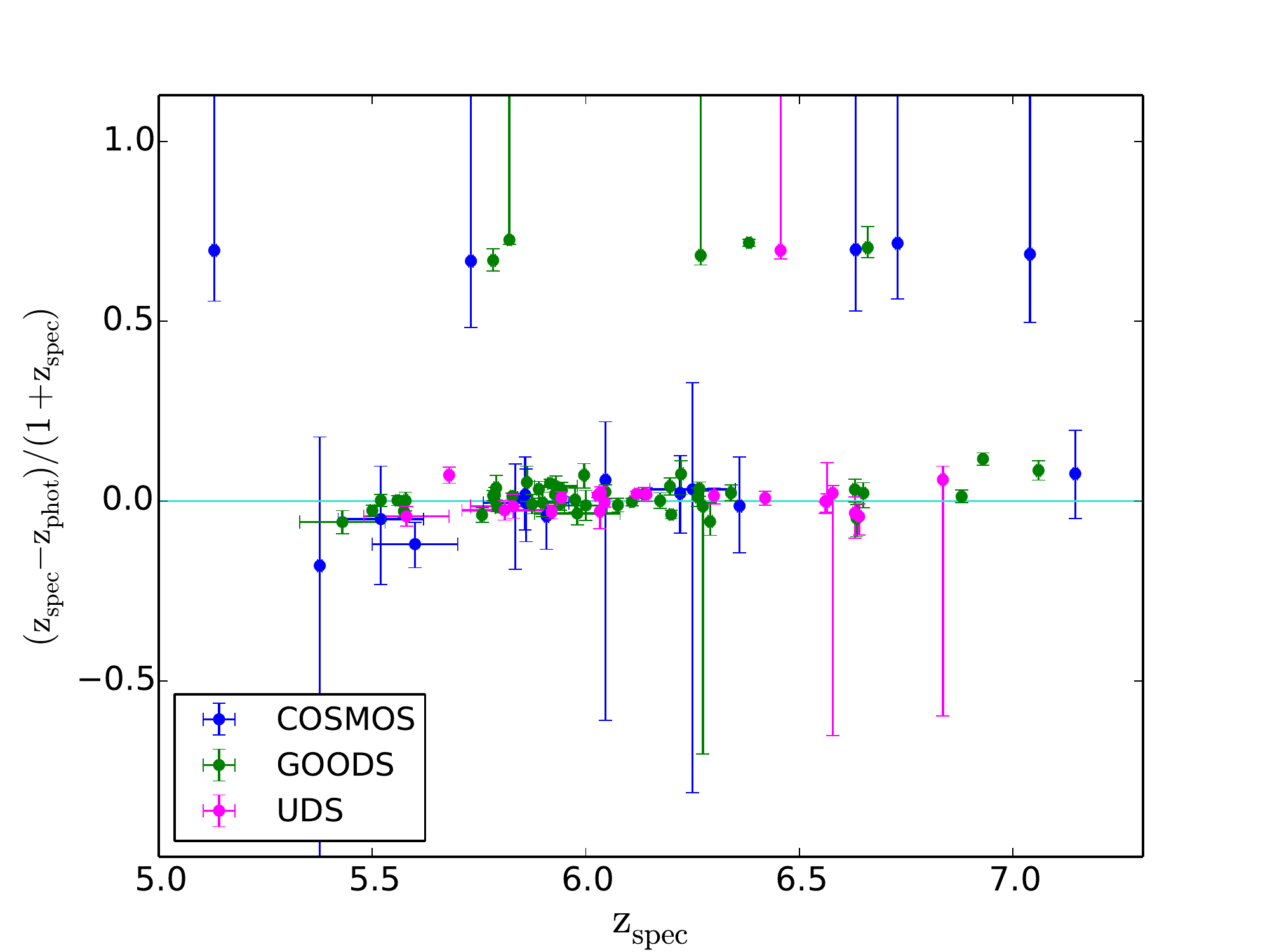}
\caption{ $\Delta z =(z_{spec} -z_{phot}) /(1+z_{spec}) $  as a function of total H-band magnitude (top) and $z_{spec}$ (bottom). The blue symbols are galaxies in the COSMOS field, the green ones are in  the GOODS-South field and the pink ones in the UDS field.  }
\label{fig:zdiff}
\end{figure}

We wish to quantify the percentage of  outliers and the accuracy of the CANDELS photometric redshift, i.e., the mean offset between  $z_{spec}$ and  $z_{phot}$   (bias) and the rms  based on our spectroscopic samples.
 To enlarge our statistics,  we include   additional CANDELS galaxies in our three fields with published spectroscopic redshift at z$\geq 5.5$ from \citet{pentericci+11},\citet{pentericci+14}, \cite{caruana+14} mostly located in the GOODS-South field. We have also checked the very recent  results from the MUSE Wide field survey (\citealt{herenz+17,caruana+18}): from the 11 published spectroscopic redshifts in the range $5.5 <z<6$,  
only 4 are detected in the CANDELS catalog, with the others being below the CANDELS detection limit. Of these,  one is a new identification (ID 7538 z=5.52027). Finally we checked the DR1 of the VUDS survey \citep{lefevre15}: of the 6 galaxies with spectroscopic redshift $>5.5$  and good quality flag ($QF \geq2$), 3 are newly identified galaxies in the COSMOS field. 
\\
 In Figure \ref{fig:specz} (left panel)  we plot the spectroscopic redshift of all these galaxies  and compare them to the photometric redshifts  obtained by the CANDELS team. The targets are color-coded depending on the field. The error bars in the photometric redshift represent the 68\% upper and lower uncertainties.  It is evident that there is a small  fraction of objects with low photometric redshift, typically around 1, but   high spectroscopic redshift: these objects were  selected from the color criteria. For most of these objects  the photometric  redshift uncertainty  at 68\% is very large and includes the high redshift solution  compatible with the real  spectroscopic redshift. The relative number of discrepant objects is higher for the  COSMOS field (blue symbols)  where the photometric redshifts are  indeed somewhat less accurate because of the fewer deep photometric bands available. In the right panel of the Figure, we show the same plot but we color-code the targets depending on the QF of the spectroscopic redshift.  We notice that most of the galaxies  with the lowest QF are at  $z<6.4$. These are typically the objects where the  redshifts are determined from the Lyman break.
 \\
In Figure \ref{fig:zdiff} we plot the $\Delta z =(z_{spec} -z_{phot}) /(1+z_{spec}) $  vs the $H_{160}$-band magnitude (upper panel) and versus $z_{spec}$ (lower panel). 
If we define the catastrophic  outliers  as  the  objects for which $\Delta z  > 0.15$ (e.g., as in \citealt{dahlenetal2013}), the fraction of such galaxies   is 14\% which is  substantially higher than what found at lower redshift for the rest of the CANDELS catalog ($\sim$3\%, see \citealt{dahlenetal2013}). The fraction of outliers is higher at fainter magnitudes and higher redshifts.
After removing the outliers, the bias, i.e., the mean of $\Delta z$ is 0.007. The bias is not constant but  depends on the redshift and magnitudes of the sources. In particular   for the most distant objects ($z>6.8$)  and for galaxies fainter than $H\sim 27$ we always find $z_{spec}  \geq z_{phot}$, with a mean bias of 0.07 and 0.04 respectively. 
The bias cannot be produced by the uncertainties in the spectroscopic redshift measurements, since  all the objects with the largest  
uncertainty are actually the brightest sources in the lowest redshift range. Since objects at $z>6.8$ are strong line emitters, one   possible  explanation is that the presence  of the line influences the determination of photometric redshifts. Similar offsets were found e.g., by   \cite{oya+16} for LAEs at $z\sim 4$  and by \cite{brinchmann+17} for galaxies with  $z>3 $ and $H<27$   in the MUSE deep field. 
After removing the catastrophic outliers the  $\sigma$ of the distribution is  0.036. Therefore  while the outliers fraction is about 3 times higher  than that of the entire CANDELS  catalog at all redshifts,  the rms is not much higher even for the most distant sources.

\subsection{\lya\ rest-frame EW distribution at z=6 and 7 }

Using the new CANDELSz7 observations and previous samples, we can now derive the \lya\ REW distributions both at z$\sim$6 and z$\sim$7  as well as the  fractions of Ly$\alpha$ emitting LBGs for faint and bright galaxies. 
Since we want to obtain results that are as robust as possible and are not subject to e.g. field to field variations, we enlarge our statistics by  including all  previous observations available, both from our group and from the literature. Clearly this means that the resulting sample is less homogeneous than CANDELSz7 alone, but this is compensated by the higher  number of galaxies observed, especially at faint magnitudes with the inclusion of lensed objects. In addition, the larger number of independent fields included mitigates uncertainties due to field-to-field variations that are very important in a partially neutral Universe (e.g., \citealt{jensen13}). In the following, we first describe in more details the sample at z=6 and z=7 that we use, we then discuss  the derivation of the REW limits for all galaxies, with the help of dedicated simulations,  and finally we derive the REW distributions and Ly$\alpha$ emitters fractions in the two redshift bins.

\subsubsection{The samples at z=6 and z=7}

At z$=$6 we use the sample that is described extensively  in  De Barros et al. (2017): briefly, it  consists of 127  galaxies selected as i-dropouts, of which 79 (>62\%) have a confirmed redshifts between 5.5 and 6.5, mostly from the \lya\ line and in few cases from the Lyman break. The galaxies come from our new program CANDELSz7 as well as previous works mostly by our team (\citealt{pentericci+14,caruana+14,pentericci+11,vanzella+11,fontana+10,vanzella+08}). Most galaxies have been selected  from the CANDELS fields (GOODS-South, UDS and COSMOS),  with a small subset coming from the NTT and BDF fields \citep{castellano+10b}.  All galaxies have been observed with FORS2, although the integration times are different and the set up of the  earlier observations by \citet{vanzella+08} had a lower resolution.

At z$=$7 we start from the sample that  we  previously assembled and analyzed in  LP14  and add galaxies  observed within  CANDELSz7 and selected with the color criteria described in Section 2. This sample consists of 134 objects, of  which 30 (22\%) have a confirmed redshift between 6.5 and 7.2  all from the \lya\ emission line. 
In this sample, galaxies come from nine  independent fields, namely  UDS, GOODS-South, COSMOS, BDF and NTT (see previous references), with the addition of few objects from the  Subaru XMM Deep Survey (SXDF) and GOODS-North from \cite{ono+12}, 10 lensed galaxies from  the observations of the Bullet cluster \citep{bradac12} and some lensed galaxies from the  Abel1703 field \citep{schenker+12}. All data with the exception of  those by \cite{ono+12} and \cite{schenker+12}, were obtained using  FORS2 with the same set-up employed for CANDELSz7, and  variable integration times ranging from a minimum of 10 to a maximum of 27 hours. 
\begin{figure*}
\includegraphics[width=9.5cm,clip=]{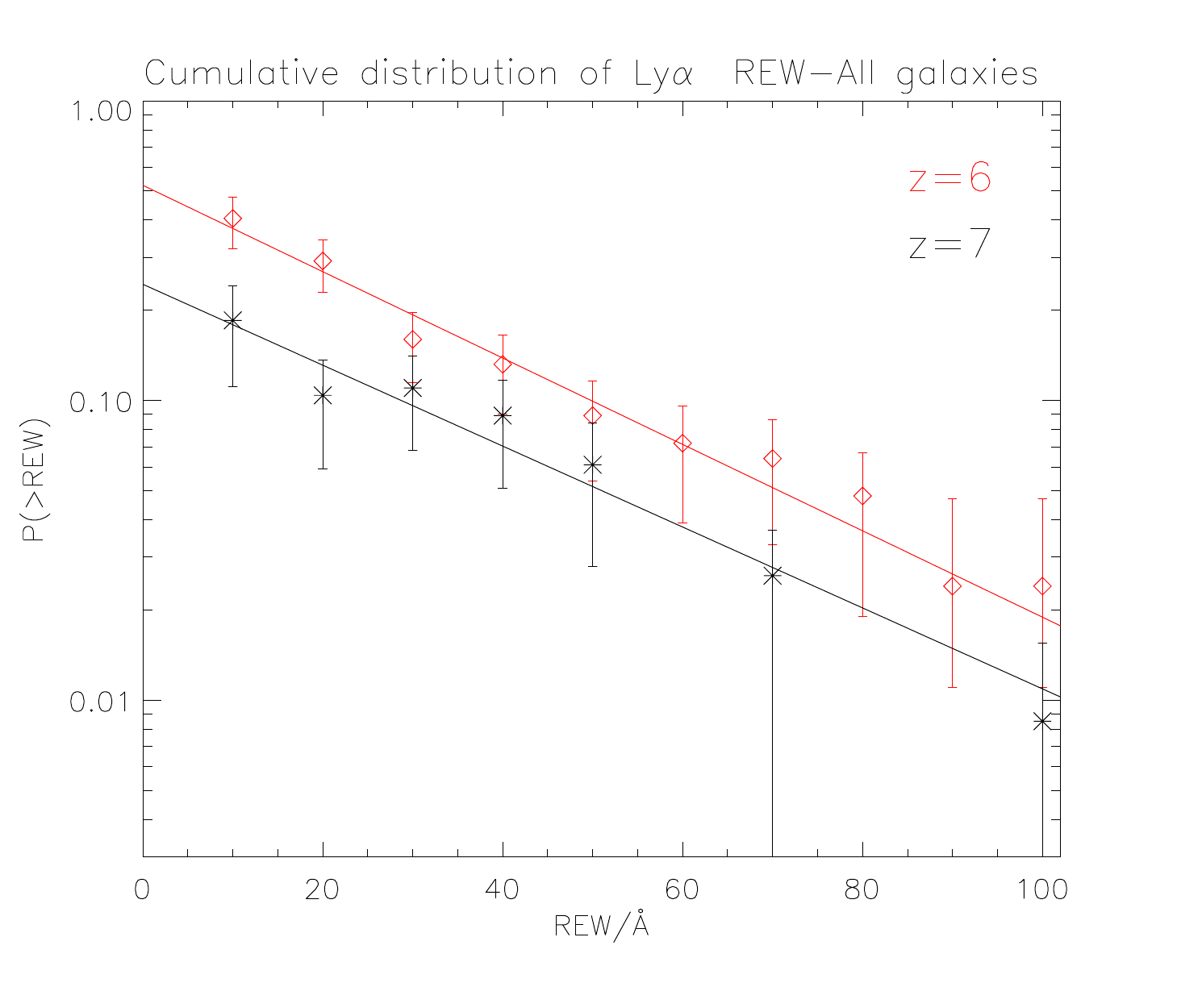}
\includegraphics[width=9.5cm,clip=]{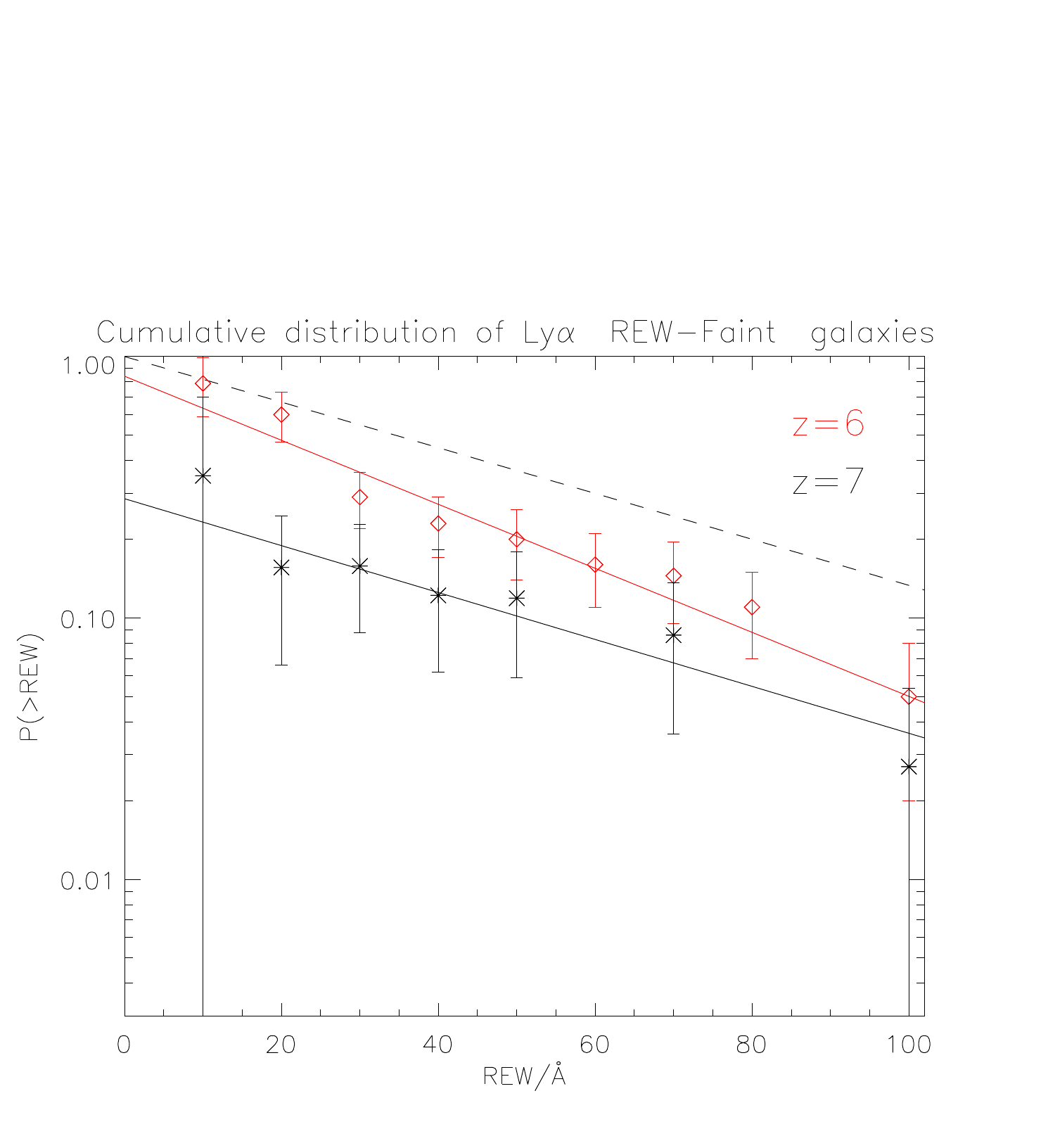}
\caption{Left: the cumulative distribution of \lya\ REW P(>REW) for the complete sample of  redshift 6 and 7 galaxies in red and black respectively. The lines are the best fir exponential functions of the two distributions. Right:  the same distribution but only for galaxies with $M_{UV}$>-20.25 (the faint sample). The dashed line indicates the previously adopted fit to the  distribution at z$\sim$6 (e.g., Dijkstra et al. 2011, LP14).  }\label{fig:dist}
\end{figure*}
\subsubsection{The fraction of Ly$\alpha$ emitters in LBGs }
To accurately derive  the fractions of Ly$\alpha$ emitters in our sample,  we must first  estimate the REW limits for each target (both detected and undetected). To do this we employ the simulations which were extensively discussed in \citet{vanzellaetal2014} and \citet{pentericci+14} and were specifically tailored for observations with the  FORS2 600z grism  (see in particular Figure 3 of \citealt{pentericci+14}) and take into account the instrument throughput and the presence and strength of the skylines.
The limit achieved on  the \lya\ line depends on the integration time, on the continuum flux of the objects and  also on the exact redshift of the line. For the objects without a confirmed redshift   we  assume z$=$6 for the i-dropouts and z$=$6.9 (the approximate mean  of the redshift PDF) for the z-dropouts.
For the  objects with a detected \lya\ emission line or with a Lyman  break, we also determine the 3$\sigma$ limit of  the line at the exact redshift position. 
For the \cite{ono+12}  sample we use the 3$\sigma$ limits on the flux reported in Table 2 of their paper, while for the 4 objects by \cite{schenker+12} we derive approximate 3$\sigma$ limits from the information provided in the paper. To convert the $3 \sigma$ limit on the emission line flux to REW limits, we used the HST photometry to determine the continuum flux.
\\
In calculating the \lya\ fractions, we must also consider that for some galaxies the redshift probability distribution extends well beyond z$\sim$ 7.3, which is the limit out to which we can detect the \lya\ emission in the   FORS2 observations. In particular the ten  Bullet cluster candidates  observed by \citet{bradac12} were selected in  a way that the probability of galaxies being at $z>7.3$  is quite high, $\sim$48\% (see Figure 5 in Hall et al. 2012). This is due to the broad J-band filter ($J_{110}$) that was available for the selection. Therefore we weighted each object in the z$\sim$7 sample  by evaluating the total probability of the galaxy  being outside the redshift range that is observable by the spectroscopic setup. In practice for most of the sub-samples this probability is negligible (see Figure 6 in \citealt{ouchi+10} for the Ono et al. sample, Figure 7 in Castellano et al. 2010a for the NTT, GOODS-South, and BDF samples), while it is $\simeq 16\%$ for the UDS and COSMOS samples (which have a tail to $z \simeq 8$, see Figure 1 in Grazian at al. 2012) and 48\% for the Bullet cluster sample. No weight was applied to the z$\sim$6 sample since the redshift distribution of these targets is entirely contained in the FORS2 observed range. 
\\
To derive the {fraction of galaxies with Ly$\alpha$ emission} at various REW limits,  we proceed as follows:  for each REW value we 
consider only those  objects that have  observations  deep enough to probe this limit, given their magnitude and redshift, and regardless if they have a detected line or not.
This means that  if an object has a detected \lya\ with $REW=50 \AA$ but its observations were  deep enough  only to probe $REW>30\AA$, this galaxy  is not considered  when evaluating the fractions with lower REW limits. 
The resulting fractions of galaxies with Ly$\alpha$ REW>25\AA,50\AA\ and 75\AA\ are presented in Table \ref{tab:frac}, for faint ($-20.25<M_{UV}<-18.75$) and bright ($-21.25<M_{UV}<-20.25$) galaxies separately, using the same magnitude bins as in LP14, and for all galaxies in the sample. The errors were evaluated using the statistics for small numbers of
events by \cite{ghere}. 

Thanks to our very large samples, not only  we can determine the fractions of emitters above these commonly used thresholds (25$\AA$,50$\AA$ and 75$\AA$),  but we can also derive accurate cumulative distributions which are presented in Figure \ref{fig:dist} for all galaxies (left panel) and for galaxies fainter than $M_{UV} =-20.25$ (right panel). 
The distribution of Ly$\alpha$ emission  at lower redshift is usually represented with an exponential function $P(>REW) \propto exp[-REW/REW_0]$  (e.g.,  \citealt{gronwalletal2007,guaita10}). In the same Figures we also plot the best fit exponential that match the observations:  these have scales $REW_0 =32\pm 8$ and $REW_0 =33\pm7$ at redshift 6 and 7, respectively
 For the faint galaxies we obtain $REW_0 =35\pm 10$ and $REW_0 =48\pm 22$ at redshift 6 and 7 respectively. 
 
\subsubsection{Comparison to previous results}
{\b At z$\sim$7 we can compare the fractions of line emitters from  the new sample to our previous derivation in LP14: the new fractions  
are slightly lower than the previous values for REW>25\AA\ and very similar, within the uncertainties, for the higher REW limits, both for faint and bright galaxies (see Table 2 of LP14). 
\\
At z$\sim$6, and as already} extensively  discussed in  \cite{debarros+17}, the fraction of \lya\ emitters that we find are considerably below previous estimates.  Specifically,  previous fractions evaluated at z$\sim$6 for  galaxies with $M_{UV} > -20.25$, were  0.54$\pm0.11$ and 0.27$\pm0.08$ respectively for EW>25\AA\ and EW>55\AA\ \citep{stark+11},  while we find  0.40$\pm0.08$ and 0.16$\pm0.05$. This discrepancy can also be seen in   Figure \ref{fig:dist} (left), where we plot as a dashed line the  representation that was employed  by  \citet{dijkstra+11} and  that  matched the previous fraction of z$\sim$6 \lya\  emitters. Our new  z$\sim$6 derivation  (red line) falls below the previous one for all values above  $REW>20\AA$.

\begin{table*}
\begin{center}
\caption{Fractions of Ly$\alpha$ emitters at z=7}
\begin{tabular}{lcccccccccc}
Mag range  & REW >25\AA & REW >50\AA & REW >75\AA  \\
\hline
-21.25$<M_{UV}$<-20.25 & 
0.09$^{+0.07}_{-0.04}$ & 
0.07$^{+0.05
}_{-0.03}$ &  <0.04\\
\\
-20.25$<M_{UV}$<-18.75 & 
0.14$^{+0.11}_{-0.07}$ & 
0.10$^{+0.09}_{-0.05}$ & 
0.06$^{0.08}_{-0.04}$ \\ \\
All  & 
0.10$^{+0.05}_{-0.03}$ & 
0.06$^{+0.03}_{-0.02}$ & 
0.02$^{+0.02}_{-0.01}$ \\
\hline
\end{tabular}
\label{tab:frac}
\end{center}
\end{table*}

We believe that the primary explanation for the discrepancy of our z$\sim$6 results with the previous derivations,  is that the  selection of our sample   is not biased by the presence of the emission line in the detection band.  Typically samples of high redshift galaxies in the pre-CANDELS epoch including i-dropouts,  were selected in the  z-band (e.g., \citealt{stark+10,vanzella+08}), which at z$\sim$6   contains  \lya. Therefore this biased positively  the fraction of strong emitters at z$\sim$6. Indeed our derivation of the \lya\ distribution at this redshift, based on an $H_{160}$-band selected sample is  lower  for objects with high REW, while it is consistent for galaxies with modest \lya\ REW (see also \citealt{debarros+17} for a more detailed discussion).   Since we have not applied any slit loss correction to our \lya\ fluxes, this could bias our \lya\ fluxes and the REW measurements to be somewhat lower than those measured in previous works. However we do not find any clear indications of slit loss corrections applied in previous works \cite{schenker+12,vanzellaetal2009}, and we remark that  the seeing conditions of our survey are excellent, given our very strict seeing limit, 
so we are confident that this is probably not the main reason for the discrepant results.

The fraction of \lya\ emitters at z$\sim$6 that we derive with the new data is similar or  even slightly lower than the fraction previously found at z$\sim$5, which was  $\sim0.48$ for the faint galaxies with REW> 25\AA\ \citep{stark+11} (see also  Figure 5 of De Barros et al. 2017).
We will discuss the redshift evolution of the \lya\ fraction more extensively in a follow-up paper. Here we just remark  that our new results indicate both a possible  flattening in  the evolution with redshift of the \lya\ fraction between z$\sim$5 and z$\sim$6,  instead of a steady increase up to z$\sim$6, and that the downturn  between z$\sim$6 and z$\sim$7 is somewhat less strong than previously reported, especially for  large values of  REW (e.g., LP14).  Assuming that the visibility of the \lya\ line depends only on the IGM neutral hydrogen content, this could mean that the increase of this quantity might be less rapid and could continue also at $z<6$. This would also  be in agreement with some recent measurements from quasar proximity zones, 
which are consistent with a  shallower evolution of the IGM neutral fraction during the epoch of reionization \citep{eilers+17}. Similarly, the recent discovery of an extreme \lya\ trough below redshift 6 \citep{becker+15} is consistent with the scenario where  reionization may be still ongoing at z$\sim$6, and be  fully completed only by z$\sim$5.5.

\subsection{The shape of the \lya\ emission}
\begin{figure}
\def\big{\includegraphics[width=10cm]{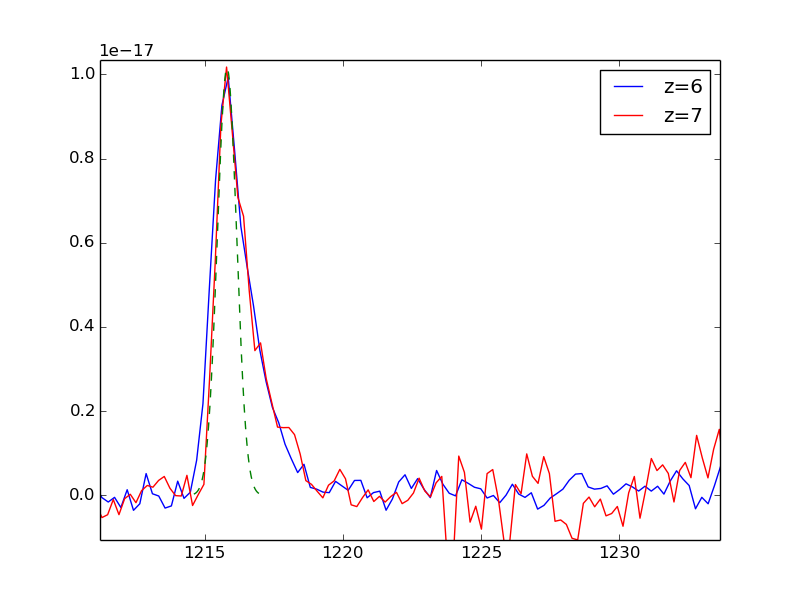}}
\def\little{\includegraphics[width=4.2cm]{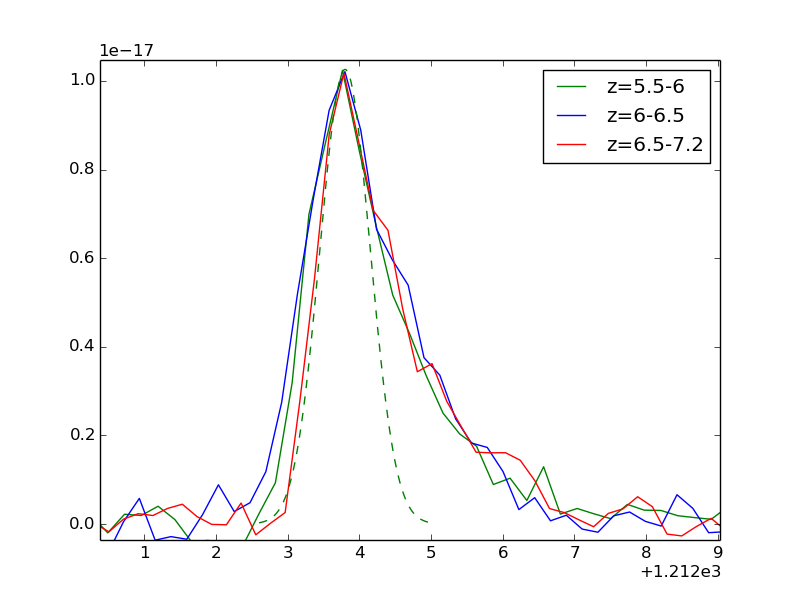}}
\def\stackalignment{l}
\topinset{\little}{\big}{2cm}{5.2cm}
\caption{The stacks of the $z\sim6$ and $z\sim7$ sources (in blue and red respectively)  showing a very asymmetric \lya\ line in both cases: the $z\sim6$ stacked line  is broader than the higher redshift line. The dashed green line is the resolution of the FORS2  spectra. Note also the positive continuum redward of the line for the $z\sim6$ stack, while in the $z\sim7$ case the flux  is consistent with the noise. In the inset,  the \lya\ line from the stacks, with the lower redshift sample split into two bins (5.5-6 in blue and 6-6.5 in green), which are consistent with each other. }\label{fig:stack}\end{figure}

\begin{figure*}
\includegraphics[width = 20cm,height=18cm,clip=]{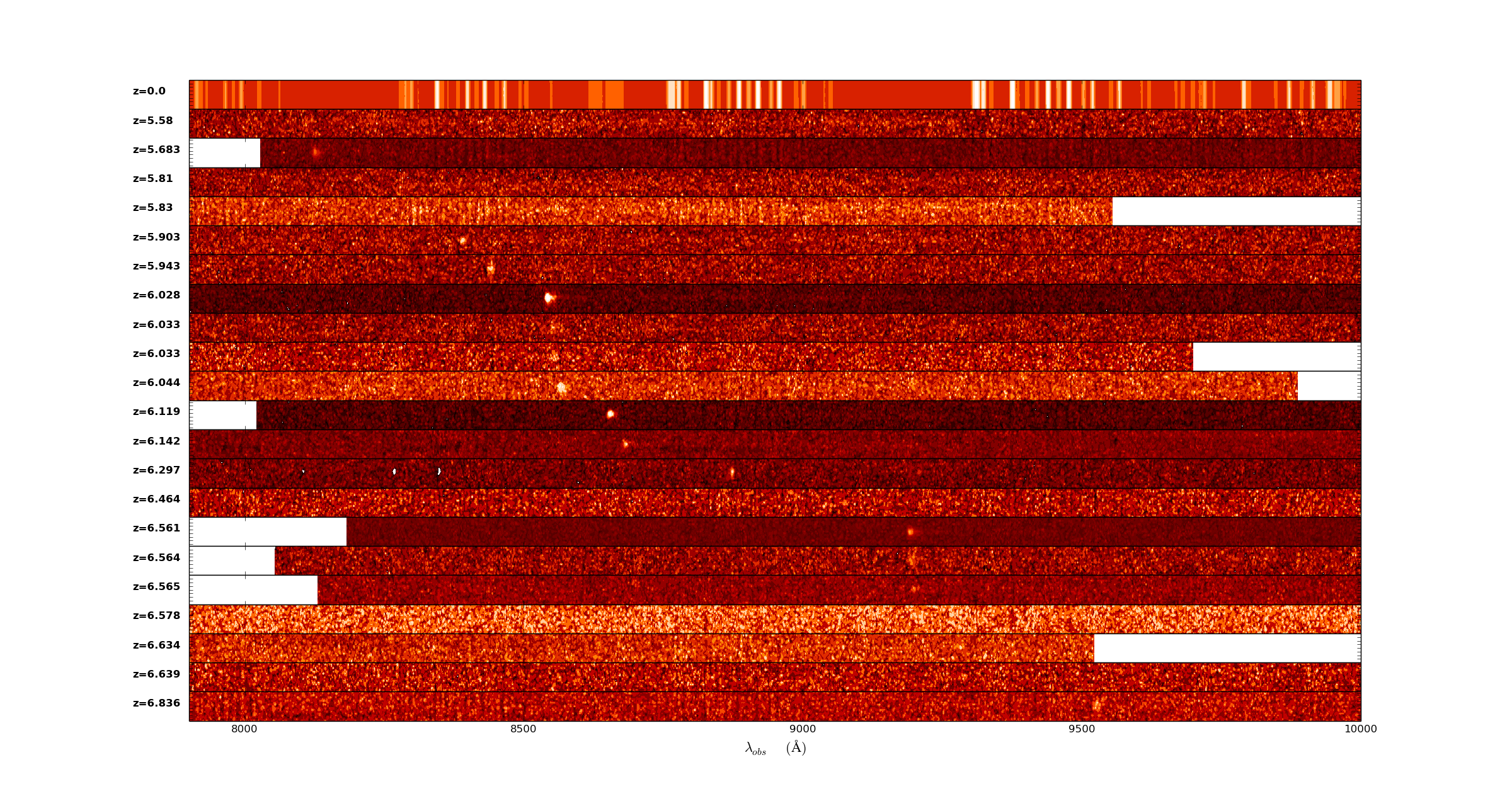}
\caption{The 2-dimensional spectra of the 21 newly confirmed galaxies in the UDS field from top to bottom in order of increasing redshifts registered to the same observed wavelength range. We plot here the maps of the S/N which were obtained from the sky subtracted data dividing them by the  map of the noise spectrum. The top 5 galaxies show a faint continuum.  The bottom panel represents the sky spectrum}
\label{fig:specUDS}
\end{figure*}

\begin{figure*}
\includegraphics[width = 21cm,height=24cm]{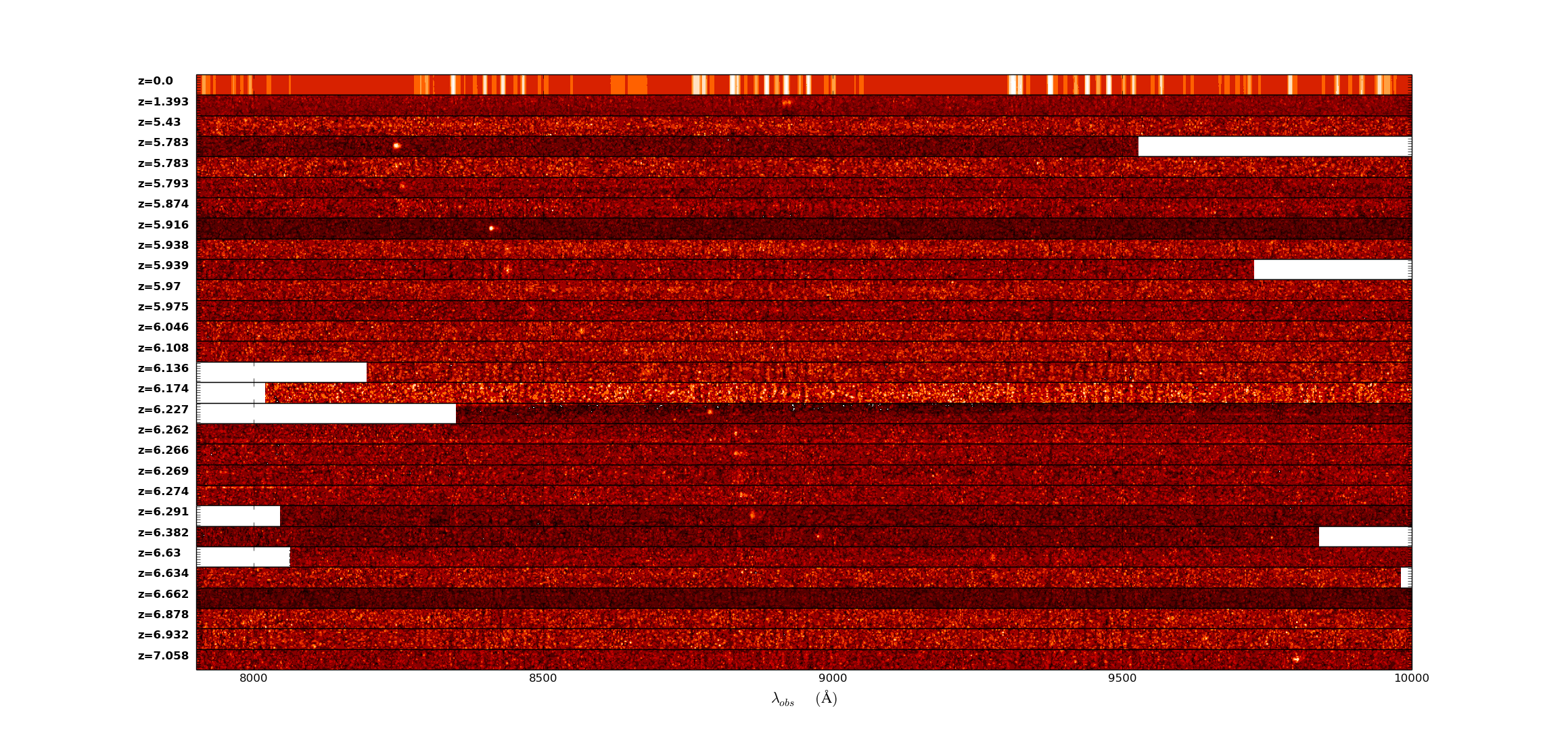}
\caption{The 2-dimensional spectra of the 28 newly confirmed galaxies in the GOODS-South  field from top to bottom in order of increasing redshift, registered to the same observed wavelength range. We plot here the maps of the S/N which were obtained from the sky subtracted data, by dividing them by the  map of the noise spectrum.   The top panel represents the sky spectrum.}
\label{fig:specGOODS}
\end{figure*}

\begin{figure*}
\includegraphics[width = 21cm,height=15cm]{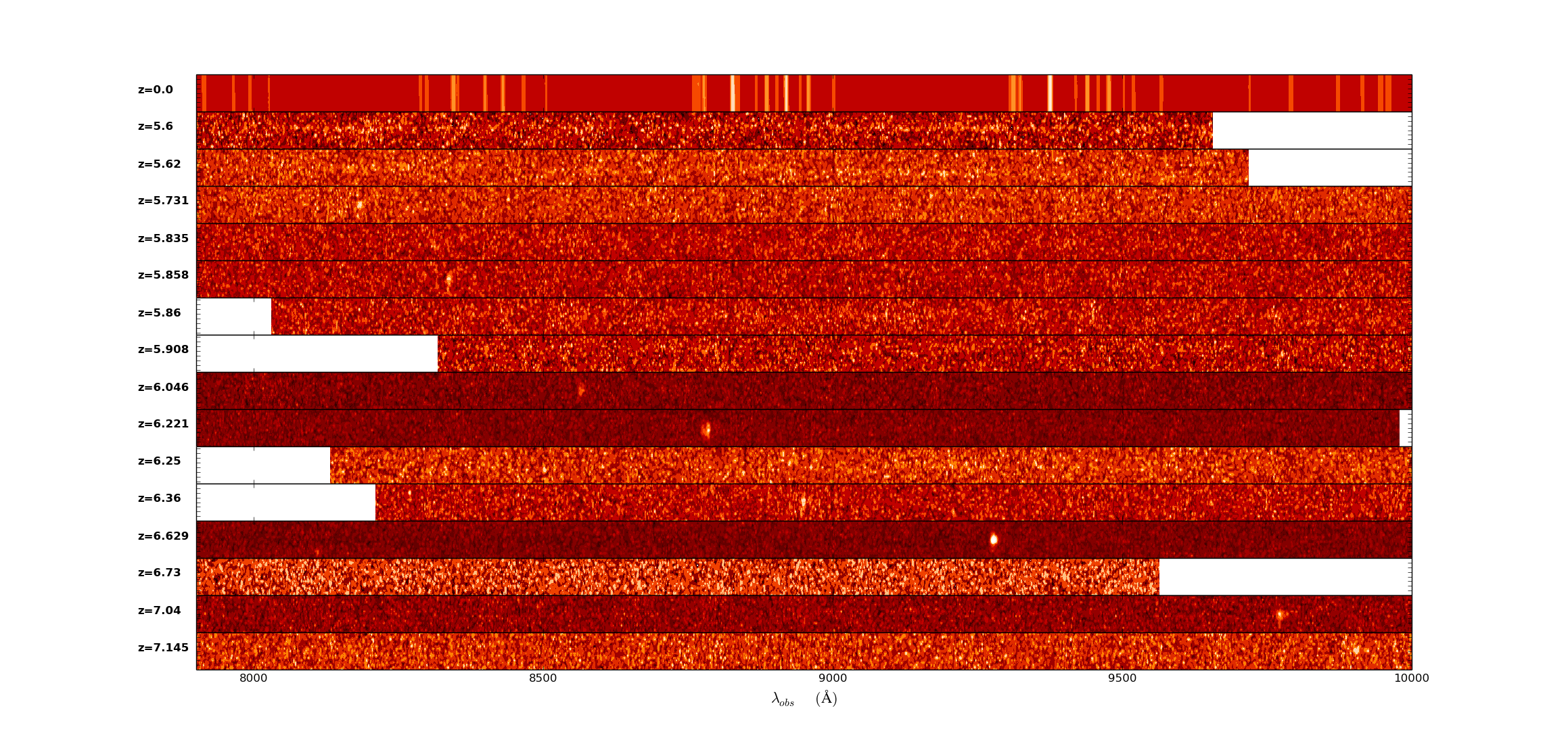}
\caption{The 2-dimensional spectra of  15  newly confirmed galaxies in the COSMOS  field, from top to bottom in order of increasing redshifts, registered to the same observed wavelength range. We plot here the maps of the S/N, which were obtained from the sky subtracted data  by dividing them by the  map of the noise spectrum.   The top panel represents the sky spectrum.}
\label{fig:specCOSMOS}
\end{figure*}

 The shape of the \lya\ line is potentially another tool to probe the reionization epoch: its width and the asymmetric properties are  expected to change in a partially neutral IGM
 (e.g., \citealt{dijkstra+07}). 
 With our medium resolution spectra (R=1390) we can investigate the evolution of the line profile of  \lya\ emitting galaxies  between z$\sim$6 and z$\sim$7. Because the S/N of individual spectra is mostly too low for an accurate spectral fit, we produced stacked spectra of all galaxies in the two redshift intervals investigated: we considered 19 galaxies at $z>6.5$ (the ones from CANDELSz7, and previous spectra from V11, LP11 and LP14) and 50 galaxies at $5.5<z<6.5$ (same references).  
To produce the stacks, we first shifted each one-dimensional spectrum to its rest-frame using the redshift evaluated from the peak of the \lya\ line. We then re-sampled each spectrum to the same grid that goes from 1100 to $1250 \AA$ with a step of $1.6\AA/(1 + z_{median}$) where  1.6\AA\ is the nominal resolution of the observed-frame spectra, using a linear interpolation. To take into account the noise of each spectrum, we computed the stack as a weighted average of the spectra, using the S/N of the \lya\ lines as weights. These have been evaluated by dividing the total flux of the \lya\ line
by the noise of the spectrum, estimated as the dispersion around the mean
value in the wavelength range $1225-1250\AA$. We chose this relatively small interval because it is the only range that is present in all spectra and that is not affected by the flux of the \lya\ emission line.
We did not normalize the spectra during the stacking
procedure, since they have basically the same level of the continuum.
The final stacks for the two  redshift bins are presented in Figure \ref{fig:stack}.  We can see that the \lya\ emission is in both cases very asymmetric with extended red wings. The blue side of the line is completely compatible with the instrumental resolution for the z$\sim$7 stack, while  it is  broader for the z$\sim$6 stack.
 From the figure  we can also see that  there is a faint continuum  in the lower redshift  stack, red-ward of the emission  line   which is not present in the higher redshift one, where the flux is consistent with zero.
\\
Since  the z$\sim$6 sample is much more numerous than the z$\sim$7 one,  part of the broadening could in principle be due to the imperfect alignment of the lines i.e., errors in the line-center determination for the low S/N lines. To check if this is the case, we further divided the lower redshift sample into two  bins, one containing galaxies in the interval z=[5.5-6], and the other containing galaxies in the interval z=[6-6.5]. In the inset of Figure \ref{fig:stack} we show that  the two lower redshift stacks are consistent with each other and that  the line is in both cases broader that at z$\sim$7, so the result is not due to the sample size. 
\\
We determine the FWHM  of the stacked \lya\ line: we first  fit a simple  Gaussian  profile to the stacks and de-convolve the values by the  resolution of the FORS2 spectra. The velocity widths we obtain are $300\pm30$ km s$^{-1}$ and $220\pm30$ km s$^{-1}$, respectively at z$\sim$6 and z$\sim$7, confirming that the line is slightly broader (at 2$\sigma$ significance) in the lower redshift stack. Since the   lines are very asymmetric,  the Gaussian fit is not a good representation of their shape. If we measure the FWHM directly from the line i.e., with no fit we find just slightly smaller values, 290$\pm 25$ km s$^{-1}$ and 215$\pm20$ km s$^{-1}$ respectively, also after deconvolution with the instrumental resolution.  In all cases the uncertainties at 68\% level  are derived using the bootstrapping statistics, by creating  100 realizations of the stacks and randomly extracting N galaxies with replacement.
 To quantify the asymmetry of the \lya\ line, 
 \cite{shima2006} introduced the weighted skewness $S_w$  which is  the skewness (or third moment) of the line multiplied by $(\lambda_{10,r} - \lambda_{10,b})$  where $\lambda_{10,r}$ and $\lambda_{10,b}$
are the wavelengths where the flux drops to 10\% of its peak value at the red and blue sides of the emission, respectively.
Since the \lya\ emission of high-redshift galaxies tends to be wider than other emission lines of nearby galaxies in the observed frame, this factor  enhances the difference between \lya\ and other lines.
Typically  large positive $S_W$ values are found for high redshift \lya\ while the  [O III] and $H\alpha$ lines are nearly symmetric, i.e., $S_W = 0$ and [O II] emitters are also expected to have small negative skewness.
 Therefore  this parameter can help  distinguish \lya\ from other emission lines. Usually the \lya\ has $S_W>3$ and this has been used as a threshold to distinguish this line from others in low resolution spectra. 
 We evaluated the skewness for the two stacked spectra and the results  are $S_W=15.8\pm8$ and $S_W=25\pm10$, respectively for the z$\sim$6 and z$\sim$7 samples, with the errors at 68\%\ evaluated using the bootstrapping statistics as above. The high redshift line is slightly more asymmetric (1$\sigma$ difference) that the low redshift one. 
% We also evaluated  the wavelength ratio (a$_\lambda$) defined by %\cite{ro2003} as 
%$a_\lambda=(\lambda_{10,r} - \lambda_p)/(\lambda_{p} - \lambda_{10,b})$, %where $\lambda_p$ is the wavelength of the peak flux density and  the %values are 2.0 and 4.3 respectively for the $z\sim6$ and $z\sim7$ stacks.
For comparison,  similar stacks made by  \citet{ouchi+10} on samples of LAEs, with 19 objects at $z=6.6$ and  11 at $z=5.7$ having \lya\ luminosities in the same range as our samples ($L=10^{42}-10^{43}$ erg s$^{-1}$), resulted in   FWHM velocity widths of $270 \pm 16$ km s$^{-1}$ and $265 \pm 37$ km s$^{-1}$ at $z=6.6$ and 5.7 respectively, with no evidence for evolution.
Similarly \cite{u+15} found no  evidence of evolution in \lya\ asymmetry or axial ratio with look-back time in high redshift LBGs,  although their sample contained only four   galaxies at $z>6$. 

As extensively discussed by \cite{dijkstra+07} the observed shape of the \lya\ line depends on many factors related to the galaxies' properties, including the star formation rate, the intrinsic width (related to the velocity dispersion of the systems) and the systemic velocity of the lines. For example galaxies with large star formation rates tend to have more symmetric lines.  In our samples, galaxies span a very similar $M_{UV}$ range (approximately between -18.5 and -22) and have similar median \lya\ luminosities although the z$\sim$6 sample  extends to  slightly higher and slightly lower luminosities than the higher redshift one (see Table \ref{tab:lya} for median values). Therefore the small differences in the observed shape of the \lya\ profile, particularly at the blue side,  might be due instead to the impact of the IGM. Our results are in qualitative agreement with the simulations by \cite{laursen+11}, who showed that at z$\sim$6 in some cases an appreciable fraction of the blue wing of the \lya\ line can still be transmitted through the IGM, especially for more massive galaxies ($M \geq 1.5\times 10^{10} M_{\odot}$ and $L_{\lya} \geq 10^{42}$ erg s$^{-1}$ ) while at $z>6.5$ the blue wing is always completely erased, regardless of the galaxies' properties.

\section{Summary and conclusions}
We have presented the results of an  ESO spectroscopic large program aimed at exploring the reionization epoch by observing a large and homogeneous sample of star forming galaxies at redshift between 5.5 and 7.2 to set  firm constraints on the evolution of the \lya\ emission fraction at this epoch.
Galaxies were selected from the $H_{160}$-band  CANDELS catalogs in the GOODS-South, UDS and COSMOS fields, using  standard color criteria and/or  the official CANDELS photometric redshifts. Spectroscopic observations of 167 high redshift galaxies were carried out with FORS2, using a medium resolution  red grating.
In 67 objects we could determine a redshift, mostly from the presence of a single  \lya\ emission line or, in few cases,  from the detection of continuum flux with a sharp drop that we interpret as the Lyman break. Two  galaxies are low redshift interlopers. Overall, the success rate for the  identification of the high redshift targets for which data could be reduced in a satisfactory way, is 40\%. Our sample increases substantially the number of sources with secure spectroscopic redshifts  in the CANDELS fields, especially at z>6.5, including 3 new galaxies at z>7.

With the newly confirmed galaxies,  as well as previous spectroscopic redshifts available  in the same fields, we  evaluated the accuracy of the CANDELS photometric redshifts at $z \geq 6$. We found that the fraction of catastrophic outliers is  14\%, i.e. more than 3 times higher than for the lower redshift galaxies in the rest of the CANDELS catalog, where it is only $\sim$3\%.  After removing the outliers, the rms uncertainty is 0.036.
We also found that photometric redshifts are in general underestimated for galaxies with $H_{160}>27$ and $z>6.8$, probably due to the presence of a strong \lya\ emission line that influences the 
broad band photometry, and which is not taken into account in the photo-z models employed.

Using our medium resolution spectra we have analysed the average shape of the \lya\ line  by creating spectral stacks in two redshift bins.  We  found that at z>6.5 the blue  side of the \lya\ emission  line is completely erased and it is consistent with the spectral resolution,  while at lower redshift a fraction of the blue wing is still transmitted. The \lya\ emission has a smaller FWHM  and is slightly more asymmetric at z$\sim$7 compared to z$\sim$6. 

Finally  we have evaluated the distribution  of the \lya\ rest-frame EW  using the new detections as well as  the accurate upper limits determined through extensive simulations, for all the objects where no emission line is observed. 
The fraction of \lya\ emitters  that we measure  at $z=6$ is consistent with previous determinations only for REW$\leq 20\AA$,  and it is below for larger REW, with a difference that reaches a factor larger than 2 at the highest REWs probed. The  fraction of \lya\ emitters at $z\sim6$ is actually  consistent with the one determined at $z\sim5$ (e.g., \citealt{stark+10}) indicating a possible flattening in  the evolution with redshift  between z$\sim$5 and z$\sim$6,  instead of a steady increase up to z$\sim$6.
The frequency of \lya\ drops by an average factor of 2 between z$\sim$6 and z$\sim$7 for galaxies with $M_{UV}>-20.25$. Overall our results might indicate a possibly slower and more extended reionization process, and
in  a future  paper we will use our new data to set more stringent  constraints on the models. In particular,  it was shown by \cite{kaki16} that improved constraints can be derived  by analyzing the full $M_{UV}$-dependent redshift evolution of the \lya\ fraction of Lyman break galaxies, such that it would be possible to distinguish between the effect of a  `bubble' model, where only diffuse H I outside ionized bubbles is present,  and  the `web' model, where H I exists only in over-dense self-shielded gas.
\section*{Acknowledgement}
This work was supported by  PRIN-INAF-2012: "From the re-jonization epoch to the 
peak of star formation: investigating galaxy mass assembly and evolution".
RM acknowledges support by the Science and Technology Facilities Council (STFC)
and the ERC Advanced Grant 695671 "QUENCH".
\begin{table*}
\begin{center}
\caption{Properties of confirmed galaxies  in the CANDELS UDS field}
\begin{tabular}{lccccccccccccc}
\hline
ID  & RA         &  Dec       &   $H_{160}$  & zphot & zspec & flux           & QF  & EW$_{Ly\alpha}$  & $\beta$ & M$_{UV}$ & $r_e$  & sel & comm.\\
             & J2000      & J2000      &           &       &       &erg/s/cm$^{2}$&     &   $\AA$  &    &       & $''$      \\
             \hline
1920 &  34.4887581  & -5.2656999  &  25.20 &  6.560 &   6.565 &  3.3e-18 &  B   &  3  & -2.95 & -21.93 & 0.12 & 1,3 &  Ly$\alpha$ \\
4812 &   34.4757347  & -5.2484999 &  25.87 &  6.566 &   6.561 &  1.9e-17 &  A   &  44 & -2.37 & -20.99 & 0.15 & 1,3&  Ly$\alpha$ \\
4872  &  34.4820328 &  -5.2481742 &  25.49 &  6.563 &   6.564 &  5.1e-18 &  A/B &  10 & -1.76 & -21.30 & 0.10 & 1,3&  Ly$\alpha$ \\
12123 &  34.4674568 & -5.2081060 & 25.03  &   1.261 &  5.903  &  6.5e-18 & A    &     & -1.09 & -21.35 & 0.14 & 2 & Ly$\alpha$ \\
%11835 & 34.253719    &  -5.2095661 &  25.88 & 6.134   &   5.92  &  00      &  C   & $<3$& -1.81 & -20.70 & 2,3  & break \\
% questo sopra lo tolgo perche era la maschera UDS0 di Adriano
14549 &  34.4828377 &  -5.1953101 &  26.75 &  6.233 &   6.033 &  5.1e-18 &  B   &  20 & -2.19 & -20.12 & 0.04  & 2,3&  Ly$\alpha$ \\
14715 &  34.4671059 &  -5.1944642 &  25.74 &  6.413 &   6.578 &  1.3e-18  &  C  &  4  & -1.21 & -20.87 & 0.25  & 1,3&  Ly$\alpha$ \\
14846 &  34.5037651 &  -5.1938372 &  25.59 &  5.912 &   6.028 &  4.8e-17  &  A   & 11 & -0.93 & -20.62 & 0.09  & 2,3 &  Ly$\alpha$ \\
14990 &  34.3546486 &  -5.1930451 &  26.54 &  6.193 &   6.297 &  7.7e-18  &  A/B & 12 & -2.84 & -20.49 & 0.19  & 2,3 & Ly$\alpha$ \\
15559 &  34.2315788 &  -5.1897931 &  26.17 &   6.073 &  6.044 &  7.9e-18  & A    & 79 & -1.98 & -21.01 & 0.19  & 2,3& Ly$\alpha$ \\
16291$^a$ &  34.3561440 & -5.1856260 & 25.87 &  6.967 &  6.638  & 1.5e-18  & B/C & 6  & -2.47 & -20.98 & 0.18  & 3 &  Ly$\alpha$   \\
18087  & 34.3972206 &  -5.1756892 &  25.65  & 5.974 &   6.119 &  3.2e-17  & A     & 47 &-2.14 & -21.22 & 0.07  & 2,3 &  Ly$\alpha$\\
18131  & 34.4512749 &   -5.1754861 & 25.493 & 5.985  & 5.81   &   00       & C   &<4  & -1.74 & -21.03 & 0.08  & 2,3   & break \\
18915 &  34.2780228 &  -5.1713920 &  25.59 &  5.859 &   5.58  &  0.0      &  C  & <1.5& -2.62 & -20.69 & 0.39  & 2,3 &  break  \\
19841 &  34.3490982 &  -5.1662178  & 26.34 &  6.370 &   6.836 &  5.0e-18 &  A/B & 18 &  -1.27 & -20.33 & 0.32  & 2,3 & Ly$\alpha$\\
23719 &  34.3104324 &  -5.1456208 &  26.79 &  5.198 &   5.683 &  9.9e1-8  &  A   & 64 & -2.36 & -20.03 & 0.17  & 2,3 & Ly$\alpha$\\
23802 &  34.2283478 &  -5.1474319 & 26.18 &  6.884 & 6.634 & 1.7e-18  &  B/C  & 7 & -2.20 & -20.63 & 0.21  & 1,3 & Ly$\alpha$\\
25826 &  34.2333316 &  -5.1364809 &  25.45  & 5.928 &   5.83 &  0.0       &  C  & <2  & -2.10 & -21.25 & 0.26  & 2,3 & break \\
28306 &  34.3560867 &  -5.2582278 &  27.13 &  6.007 &   6.142 &  6.5e-18  &  A  & 41  & -2.74 & -19.99 & 0.05  & 2,3 &   Ly$\alpha$\\
29191 &  34.5253906 &  -5.2412128 &  26.48 &  5.879 &   5.943 &  8.8e-18  &  A/B & 34 & -3.18 & -20.57 & 0.06  & 2,3 &  Ly$\alpha$ \\
31124 &  34.2537117 &  -5.2067928 &  26.89 &  1.260 &   6.464 &  1.5e-18 &  B/C  & 8  & -2.06 & -19.98 & 0.15  & 2 &  Ly$\alpha$\\
33304 &  34.5186958 &  -5.1694598 &  26.26 &  5.866 &   6.033 &  3.5e-18  &  B  & 16  & -2.14 & -20.44 & 0.08  & 2,3 &   Ly$\alpha$\\
\end{tabular}
\end{center}
$^a$ This galaxy  has [CII] 158$\mu m$ detection  published in Pentericci et al. (2016).
\label{tab:UDS}
\end{table*}
\begin{table*}
\begin{center}
\caption{Properties of confirmed galaxies in the CANDELS COSMOS field}
\begin{tabular}{lcccccccccccccc}
\hline
ID  & RA         &  Dec       &   $H_{160}$  & zphot & zspec & flux           & QF  & EW$_{Ly\alpha}$  & $\beta$ & M$_{UV}$ & $r_e$  & sel & comm.\\
             & J2000      & J2000      &           &       &       &erg/s/cm$^{2}$&     &   $\AA$  &    &       & $''$      \\
             \hline
6822  & 150.177171   & 2.261854  &  25.79  & 0.848   & 5.131  &  0.0      &  B    & <19 & -1.69 & -20.50 & 0.23 & 2 & Ly$\alpha$, break  \\
7499  & 150.089164   & 2.26949   &  26.29  & 5.737  &  5.858  &  6.1      &  A/B  & 19  & -1.30 & -20.33 & 0.09 & 2,3 & Ly$\alpha$  \\
7692  & 150.107757   & 2.271918  &  26.04  & 5.634  &  6.046  &  7.0e-18  &  A    & 24  & -0.74 & -20.20 & 0.19 & 2,3 & Ly$\alpha$   \\
8118  & 150.154944  &  2.277192  &  26.07  & 1.24   &  5.731 &   2.7e-18  &  A/B  & 7   & -1.71 & -20.42 & 0.14 & 2   &  Ly$\alpha$    \\
8474 &  150.106609  &  2.281278  &  25.00  & 6.079  &  0.661 &   --       &  A    & --  & --    & --     & 0.55 & 2,3 & H$\beta$ OIII  \\
10699 & 150.118281  &  2.307277  &  26.61  & 5.778  &  5.835 &   2.5e-18  &  C    & 9   & -2.79 & -20.33 & 0.02 & 2,3 & Ly$\alpha$   \\
12306 & 150.127459  &  2.326425  &  24.31  & 6.197  &  5.908 &   2.8e-18  &  A/B  & 3   & -1.70 & -22.26 & 0.45 & 2,3 & Ly$\alpha$    \\
13679$^a$ & 150.099037 & 2.343627 & 25.43  & 6.525  &  7.145 &   9.2e-18  &  A/B & 15   & -1.54 & -21.46 & 0.01 & 1,3& Ly$\alpha$ \\
18472 & 150.126605  &  2.401444   & 25.63  & 5.899  &  5.86  &   0.0      &  C   & <5   & -1.94 & -21.03 & 0.02 & 2,3 & break  \\
20521 & 150.139594  &  2.426985  &  25.69  & 6.46   &  6.360   &  6.5e-18 &  B   &  10  & -2.17 & -21.13 & 0.17 & 2,3 & Ly$\alpha$    \\
21151 & 150.165997  &  2.435793  &  26.23  & 1.522  &  7.040   &  1.65e-17 & A   & 65   &-1.25  & -20.57 & 0.05 & 1 & Ly$\alpha$   \\
21411 & 150.183018  &  2.439205  &  25.77   & 6.066  &  6.221  &   3.8e-17  &  A  & 90   & -1.40 & -20.77 & 1.51 & 2,3 & Ly$\alpha$    \\
22592 & 150.196765  &  2.454824  &  24.15  & 6.39   &  5.60  &   0.0      &    C & <1.2 & -2.28 & -22.54 & 0.02 & 2 & break     \\
24108$^a$ & 150.197222  & 2.478651 & 25.14  & 1.297 &  6.629  &  2.0e-17 &    A  & 27   & -1.76 & -21.67 & 0.27 & 2 & Ly$\alpha$    \\
25022 & 150.191343  &  2.492327  &  25.99  & 5.847  &  5.62   &  0.0     &   C   & <2   & -1.56 & -20.37 & 0.11 & 3 &  break     \\
26366 & 150.161146   & 2.511134  &  25.84  & 6.013   & 6.25 &   0.0   &     C   & <2    & -0.89 & -20.45 & 0.04 & 3 & break      \\
30549 & 150.162535   & 2.234355   & 26.65  & 1.19   &  6.730   &  3.0e-18    &     B/C &  16  & -1.76 & -19.96 & 0.05 & 1 &  Ly$\alpha$   \\
36393 & 150.118121   & 2.451689   & 26.17   & 6.522  &  5.377   & 4.0e-17  &  B   & 12   & -2.33 & -20.146    & 0.30 & 2,3 & Ly$\alpha$        \\

\end{tabular}
\label{tab:COSMOS}
\end{center}
$^a$ These galaxies have [CII] 158$\mu m$ detections  published in \cite{pentericci+16}.
\end{table*}

\begin{table*}
\begin{center}
\caption{Properties of confirmed galaxies  in the CANDELS GOODS-South field}
\begin{tabular}{lccccccccccccc}
\hline
ID  & RA         &  Dec       &   $H_{160}$  & zphot & zspec & flux           & QF  & EW$_{Ly\alpha}$  & $\beta$ & M$_{UV}$ & $r_e$  & sel & comm.\\
             & J2000      & J2000      &           &       &       &erg/s/cm$^{2}$&     &   $\AA$  &    &       & $''$      \\
             \hline
10219 &  53.2020264 & -27.8163528 & 26.14  & 5.997 & 6.136   & 4.0e-18    & B   & 14 & -1.57 &  -20.56 & 0.22   & 2,3 & Ly$\alpha$  \\
11464 &  53.1174545 & -27.8051872 & 26.04  &  5.994 & 5.939  & 7.0e-18    & B   & 22 & -1.59 &  -20.53 & 0.14  & 2,3 & Ly$\alpha$ \\
12881 & 53.0694046  &  -27.7943935& 26.11  &  5.9360& 5.874  & 2.0e-18  & B/C & 5  & -2.22 &  -20.70 &   0.37  & 2,3 & Ly$\alpha$ \\
13065 &  53.1438484 & -27.7930164 & 28.21  &  6.001 & 6.932  & 2.2e-18  & B   & 42 & -1.90 & -18.85  &  0.02 & 2,3 & Ly$\alpha$  \\
14439 &  53.1248894 & -27.7841072 & 27.18  &  5.680 & 5.783  & 1.0e-17  &  A  & 50 & -2.86 &  -19.93 &  0.15 &2,3 & Ly$\alpha$   \\
15178 &  53.0558853 & -27.7795563 & 26.78  &  6.018 & 6.266  & 9.1e-18  & A/B & 48 & -1.58 &  -19.88 &  0.09& 2,3 & Ly$\alpha$  \\
15404 &  53.0338783 & -27.7780075 & 26.70  &  6.706 & 6.291  & 9.0e-18  & A   & 44 & -1.30 &  -19.94 &  0.09 & 2,3 &  Ly$\alpha$  \\
15951 &  53.1050529 & -27.7740688 & 26.91  &  6.997 & 6.634  & 2.0e-18  & B   & 17 & -2.03 & -19.98  &  0.08 & 1,2 &   Ly$\alpha$   \\
16024 & 53.0731506  & -27.773634  & 25.45  & 5.805  & 5.43  & 00       &  C  & <1.6 & -1.33 & -20.82&0.22 & 2,3& break \\
17692 &  53.1627693 & -27.7607594 & 28.09  & 5.574  & 5.916  & 2.25e-17 &  A  &153 &  -2.44 & -18.93 & 0.04 & 2,3 & Ly$\alpha$\\
%18694 & 53.050354   & -27.7521305 & 24.93  &        & 6.52   &  1.5e-18 & B/C & 3   &  0.15 &  -21.33 & ? & Ly$\alpha$ \\
% questo ha una riga ma non credo che sia ad alto redshift non era selezionato in nessun modo se non agn a redshift intermedio 
20698 &  53.2031670 & -27.7337036 & 25.99  &  6.168 & 6.174  &  1.5e-18 & B/C   & 4   & -1.45 &  -20.75 &0.18 & 2,3 & Ly$\alpha$  \\
22647 &  53.0666809 & -27.7170753 & 25.91  & 1.246  & 5.783  &  00      & B/C   & <11 & -1.77 & -20.78  &0.92 & 2,3 & Ly$\alpha$ \\
26628 &  53.0707664 & -27.7066383 & 26.58  & 5.949  & 5.975  &  2.2e-18 & B   & 9   & -2.47 & -20.30  &  0.29 & 2,3 & Ly$\alpha$ \\  
31131 &  53.1255341 & -27.7866764 & 26.80  & 6.00   & 6.382  & 3.1e-18  & B   & 26  & -1.0  & -19.89  &  0.42 & 2,3 & Ly$\alpha$ \\
31144 &  53.0830650 & -27.7862740 & 27.31  & 1.306  & 6.269  &  6.0e-18   & B/C & 65& -1.61 & -19.28  & 0.20 & 1   &  Ly$\alpha$ \\
32252 &  53.0951385 & -27.7605419 & 27.55  &  6.183 &  6.262 & 00       & B/C & <23 & -2.99 & -19.64  & 0.08 & 2,3 &  Ly$\alpha$\\ 
33418 &  53.1944962 & -27.7263489 & 27.40  &  6.373 & 7.058  & 1.2 e-17  &  A  & 110 & -2.2- & -19.54 &0.04 & 1,2 & Ly$\alpha$  \\
33477 &  53.0641861 & -27.7246075 & 27.51  &  6.379 & 6.274  & 6.2e-18  & A/B & 64  & -2.18 & -19.26 & 0.72 & 3  &  Ly$\alpha$  \\
34061 &  53.0461273 & -27.7082958 & 26.53  &  5.68  & 6.227  &6.8e-18  &  A  & 36  & -0.35 & -19.79 &0.18$^b$& 2,3 &  Ly$\alpha$\\
H2525$^a$ & 53.145531 & -27.783724  & 27.98 &  6.78  &  6.878  & 2.0e-18  &  B/C  & 18  & -1.98 & -19.66 &-- & 1,2 &  Ly$\alpha$\\
13184  & 53.1519318  & -27.7923527 & 27.46  & 1.264 &  6.662 & 5.0e-18    & A/B  & 81  & -0.35 & -19.05 & 0.21 & 1 &  Ly$\alpha$   \\
% 13184 e[ osservato sia da noi che da Bunker 
15443  & 53.1519432  & -27.7781773 & 25.73 & 5.933  & 5.938  &  3.6e-18  &  A   & 7  & -1.89 & -20.98  &0.15&  2,3 & Ly$\alpha$    \\
16371  & 53.1595078 &  -27.7714462 & 26.35 & 6.123  & 6.108 &  3.0e-18    &  B & 11 & -2.01 & -20.43 & 0.04 & 2,3 & Ly$\alpha$   \\
%17720  & 53.1225471  & -27.7604961 & 26.78  & 5.801 & 5.928 &  5.5e-18  & B   &  30 & -2.13 & -19.97 & 2,3 & Ly$\alpha$  \\
%questa sopra e'  gia in pentericci et al 2011 perche osservato da fontana 2010 
18310 &  53.1419334  & -27.7551537 & 27.26  & 5.867 & 6.046 & 3.0e-18    & B   & 26  & -1.66 & -19.34 &0.03 & 2,3 & Ly$\alpha$     \\
%GOODS19377 &  53.155757 -27.746067  bunker33top pubblicato in caruna e al come 237384457  \\
%GOODS20776 &  bunker24TOP\\ NON vero
22194 &  53.1161804  & -27.7210217 &  26.70   & 5.541 &  5.793  &  5.0e-18& B & 24  & -2.00 & -19.90 & 0.35& 2,3 & Ly$\alpha$     \\
26560 &  53.1581459  & -27.7021046  & 24.55  &  6.22   & 5.97  &  00    & C & <1.3 & -1.38 & -21.99 & 0.24 &  2,3 & break  \\
31759 &  53.1348877  & -27.7726326 &  27.47  &  6.174 &  1.393  &  1.27e-17 & A  & - & - & - &0.02 & 2,3 & [OII]   \\
31891 &  53.1742516  & -27.7697887  & 28.75  &  6.386  & 6.630  &  2.0e-18 &  A & 73  &-1.85 & -18.08 &0.03 & 1,3 & Ly$\alpha$    \\
%32453 &  53.17990875 & -27.7548732  & 27.08  &  5.94   &  5.93 & 8.4e-18 & A & 56 & -1.92 & -19.61 & 2,3 & Ly$\alpha$ caruana et al. 24131845175\\
%GOODS34271 bunker 6TOP\\ gia pubblicato  in pentericci 2014  
\end{tabular}
\end{center}
$^a$ This galaxy does not have a CANDELS ID, since  it is detected in the hot-mode only but it is located within the Kron radius of 
 a cold source  and therefore  excluded from the final catalog \citep{guo+13}.$^b$ The $r_e$ reported is measured in $Y_{098}$ filter since the  $J_{125}$ is contaminated. 
\label{tab:GOODS}
\end{table*}

\begin{table*}
\begin{center}
\caption{Fitted  parameters for the \lya\ emission line in the two stacks.}
\begin{tabular}{lccccccc}
$<z>$  &N &  $<L_{\lya}>$ &$ <M_{UV}>$ & FWHM    & $S_W$   \\
       &  &  erg $s^{-1}$ & &  km $s^{-1}$  \\
\hline
6.0  & 52 & $2.5\times 10^{42}$ & -20.53   & $300\pm30$     & $15.8 \pm 8$ \\
6.9  & 19 & $2.7\times 10^{42}$ & -20.62   & $220\pm25$     &  $25 \pm 10$   \\
\end{tabular}
\end{center}
$<z>$ median redshift of the sample;  N total number of objects in each sample; $<L_{\lya}>$  median \lya\ luminosity of the sample;  $ <M_{UV}>$ median $M_{UV}$; FWHM of the stacked lines from the Gaussian fit;  $S_W$  weighted skeweness parameter of the stacked lines. 
\label{tab:lya}
\end{table*}
%\begin{figure*}
%\includegraphics[width = 12cm,clip=]{/Users/laurapentericci/Documents/z7paper/24108_1d2d.png}
%\includegraphics[width = 10cm,clip=]{/Users/laurapentericci/Documents/z7paper/13679.png}
%\includegraphics[width = 10cm,clip=]{/Users/laurapentericci/Documents/z7paper/11464.png}
%\caption{The 1dimensional (top) and 2-dimensional spectra of source COSMOS24108 at z=6.632 showing a bright asymmetric Ly$\alpha$ emission line. Note that in the 2-dimensional spectrum and other emission line from a lower redshift galaxy is also visible. The 1-dimensional spectrum is smoothed by 3 pixels and some of the brightest spikes from residuals of background subtraction were removed.  }
%\label{specz}
%\end{figure*}

\bibliographystyle{aa}
\bibliography{pentericci}

\end{document}